\documentclass[prc,twocolumn,showpacs,preprintnumbers,amsmath,amssymb]{revtex4-1}

\usepackage[utf8]{inputenc}
\usepackage{graphicx}
\usepackage{float}

\usepackage{amsmath}
\makeatletter
\g@addto@macro\normalsize{%
  \setlength\abovedisplayskip{6pt}
  \setlength\belowdisplayskip{6pt}
  \setlength\abovedisplayshortskip{6pt}
  \setlength\belowdisplayshortskip{6pt}
}
\makeatother

\usepackage{xcolor}
\definecolor{blue}{RGB}{0,0,0}
\definecolor{red}{RGB}{0,0,0}

\allowdisplaybreaks
\usepackage{color}

\usepackage{multirow}
\usepackage{dcolumn}
\newcolumntype{d}[1]{D{.}{.}{#1}}

\begin{document}
\title{Photoproduction of $K^{+}\Lambda$ with a Regge-plus-resonance model}
\author{P.~Byd\v{z}ovsk\'y$^{1}$, D.~Skoupil$^{1,2}$} 
\affiliation{$^{1}$Nuclear Physics Institute, CAS, 25068 \v{R}e\v{z}, Czech Republic,\\
$^{2}$Faculty of Science, Kyushu University, Fukuoka, 819‐0385, Japan}

\date{\today }
\begin{abstract}
Making use of the hybrid Regge-plus-resonance model, we investigate the process of kaon photoproduction off the proton target. We present a new model whose free parameters were adjusted to data in and above the resonance region and which provides an acceptable description of experimental data. The overwhelming majority of nucleon resonances selected in this analysis overlaps with those selected in our previous analyses and also with the Bayesian analysis with the Regge-plus-resonance model, which we deem to be dependable. A novel feature of our model consists in applying a different scheme for gauge-invariance restoration, which results in a need for implementing a contact current. As we further reveal, the choice of the gauge invariance restoration scheme as well as the choice of either pseudoscalar or pseudovector coupling in the strong vertex play a significant role for cross-section predictions at forward angles where data are scarce. 

\end{abstract}


\maketitle

\section{Introduction}
The main objects of exploration of kaon photo- and electroproduction from nucleons are the investigation of baryon resonance spectrum and interactions in systems where hyperons and kaons arise. It can also shed some new light on an interesting topic of ``missing" resonances that have been predicted by quark models~\cite{Capstick,Loring} but have not been seen in the pion production or $\pi N$-scattering processes. What is more, an accurate depiction of the elementary production process is a necessary step for a further work on computing cross sections and excitation spectra of $\Lambda$ hypernuclei production~\cite{hypernucleus}.

There are distinct methods of describing the elementary process of photo- and electroproduction. There are on the one hand models based on quark degrees of freedom~\cite{ZpL95,LLP95,FHZ91} which introduce a relatively small number of parameters and explicitly work with an inner structure of baryons. On the other hand, we can assume hadrons as effective degrees of freedom and base our calculations on effective Lagrangians. Since in these models there is no explicit connection to QCD, the number of parameters is directly related to the number of resonances introduced. As the kaon production occurs in the third-resonance region, where many states possibly couple to the $KY$ channels, the number of resonances and consequently the number of parameters can be relatively  high. In this hadrodynamical approach, one can either assume coupling of the production channels by the meson-baryon interaction~\cite{Giessen,JDiaz,Anisovich,Borasoy} or opt for a considerable simplification which stems from neglecting the rescattering effects in the formalism and assuming that their influence is, at least to some extent, included in the effective values of coupling parameters adjusted to experimental data. Recently, we have published two analyses of $K^+\Lambda$ production with such an approach~\cite{SB16,SB18} and this framework is now also available for online calculations~\cite{IMweb}.

A significant reduction of the number of parameters can be accomplished by using Regge-plus-resonance model (RPR) constructed by the group in Ghent~\cite{RPR,RPR07,RPR11}. This model allows us to describe the kaon-hyperon photo- and electroproduction from the threshold energy up to energies well beyond the resonance region, as it is a hybrid between the isobar model suitable for calculations in the resonance region and the Regge model~\cite{Guidal} which is applicable above the resonance region $E_\gamma > 3\,\text{GeV}$. The Regge part of the amplitude, being a smooth function of energy, forms the background in the resonance region and dominates the predictions above the resonance region. On top of this Regge-like background, there are contributions of nucleon resonances added which then model the resonance part of the amplitude in the resonance region and vanish beyond it.

An important and often discussed issue of the Regge-type approach to photo- and electroproduction is the gauge invariance restoration method. A frequently used method is the one introduced by Guidal \emph{et al.}~\cite{Guidal}. They added the proton exchange with the vector photon-proton coupling to the kaon exchange contribution to construct the residual function of the kaon-trajectory part of the amplitude. 
Here we utilize a prescription suggested by Haberzettl \emph{et al.}~\cite{HH} which is based on the generalized Ward-Takahashi identities introducing a contact term. We also 
address the issue of the proton-kaon-Lambda coupling  assuming both the  pseudoscalar and pseudovector forms that influence the gauge restoration method.

In this work, we present a new Regge-plus-resonance model for production of $K^+\Lambda$ with a special emphasis on the subject of gauge-invariance restoration. As well as in our studies of $K^+\Lambda$ production with help of isobar model~\cite{SB16,SB18}, we use a consistent formalism for description of high-spin nucleon resonances~\cite{Pascalutsa,Vrancx} and pay close attention to observables predictions at small kaon angles. The latter is vital for getting reliable predictions of hypernucleus production cross sections~\cite{hypernucleus}.

This article is organized as follows. In Sec. II the frame for Regge description of the non resonant part of the photoproduction amplitude is given discussing in detail the gauge-invariance restoration method. In that part, we also introduce a novel feature of the model: the contact term. The resonant part of the amplitude is described in Sec. III. For more details on formalism of $K^+\Lambda$ photoproduction we refer to Ref.~\cite{SB16}. A method of adjusting free parameters of the model is described in Sec. IV.
The Sec. V is devoted to comparison and discussion  of model predictions with data and with results of other models. Conclusions are given in Sec. VI. 
More details on the Regge formalism are given in Appendices A and B.

\section{Regge Model}
\label{sec:Regge}
At the basis of the Regge theory~\cite{Collins} is the fact that, at energies where individual resonances can no longer be distinguished, the exchange of entire Regge trajectories predominates the reaction dynamics rather than the exchange of individual particles. This high-energy framework applies to the ``Regge limit'' of extreme forward (in the case of the $t$-channel exchange) or backward (for the $u$-channel exchange) scattering angles, corresponding to small $|t|$ or $|u|$, respectively. Since the lightest hyperon, the $\Lambda$ hyperon, is significantly heavier than a kaon and, therefore, the $u$-channel poles are located much further from the backward-angle kinematical region than the $t$-channel poles are from the forward-angle region, the $u$-channel exchange Reggeization, \emph{i.e.} the procedure of requiring the Regge propagator to reduce to the Feynman one at the closest crossed-channel pole, might not lead to good results \cite{RPR}. What is more, the high-energy data in the backward-angle region are scarce. Therefore, we have chosen to deal with the $t$-channel exchange Reggeization only.

Since in the vicinity of the $t$-channel pole the Regge amplitude is assumed to be identical with the Feynman amplitude for the exchange of the given particle, the Regge theory, in its simplest form, can be formulated by modifying the isobar model. The process of Reggeization is quite straightforward and goes as follows: One writes the amplitude for the exchange of the given particle (in the corresponding pole both Feynman and Regge propagators coincide) and then interchanges the Feynman propagator with the Regge one,
$$\frac{1}{t-m_X^2}\rightarrow \mathcal{P}_{Regge}^{X}(s,\alpha_X(t)),$$
and the remnant terms in the amplitude then labels as a Feynman residuum $\beta_X$. The amplitude constructed in this way includes effectively exchanges of all particles represented by the given trajectory and reads
\begin{equation}
\mathbb{M}_{Regge}^X(s,t)=\beta_X(t)\,\mathcal{P}_{Regge}^{X}(s,\alpha_X(t)).
\label{eq:Regge}
\end{equation}
In the case of $K^+\Lambda$ production, we Reggeize contributions of the $K^+(494)$ and $K^{*}(892)$ amplitudes only. For more details on Regge trajectories and propagators see Appendix \ref{App:Regge}.

The main asset of the Regge model is a reduced number of free parameters to be adjusted to experimental data. If we do not consider the hadron form factor for the proton exchange and terms proportional to the function $\hat{A}(s,t,u)$ in the transversal part of the contact current (see Eqs.(\ref{eq:gce}) and (\ref{eq:gce-pv}) below), then there are only three free parameters: $g_{K\Lambda N}$ and $G_{K^{*}}^{(v,t)}$.

\subsection{Gauge invariance in Regge model}
As in the $\gamma(k)+p(p)\rightarrow K^+(p_K)+\Lambda(p_\Lambda)$ reaction only the incoming proton and the outgoing kaon carry electric charge, the relevant contributions in view of gauge invariance stem from the $s$- and $t$-channel Born terms and a contact term.
In the following we use  the method of repairing gauge invariance broken by Reggeization of the $t$-channel exchanges \textcolor{blue}{and by inclusion of hadron form factors}, which was suggested by Haberzettl {\it et al.}~\cite{HH}.

The total gauge-invariant Regge current in our approach reads
\begin{eqnarray}
\mathbb{M}^\mu &=&  \mathbb{M}^\mu_{R,t} + \mathbb{M}^\mu_s + \mathbb{M}^\mu_{int} \nonumber\\
&=& J_K^\mu(p_K,p_K-k)\,\Delta_K(p_K-k)\,F_{R,t} \nonumber\\
&& \;+ F_s\,S_p(p+k)\,  J_p^\mu(p+k,p) + \mathbb{M}_{int}^\mu\,,
\end{eqnarray}
where $J^\mu$ are electromagnetic currents, $\Delta_K$ is a kaon propagator, $S_p$ is a proton propagator, and $F_{R,t}$ and $F_s$ are hadron vertices in the $t$ and $s$ channel, respectively. The contact current is given by~\cite{HH}
\begin{equation}
\mathbb{M}_{int}^{\mu}=m_c^\mu \mathcal{F}_t(t,s)+ G C^\mu,
\label{eq:Mc}
\end{equation}
where the hadron form factor $f_t(t)$ appearing in Eq. (A1) in Ref.~\cite{HH} was interchanged for the Regge residual function $\mathcal{F}_t(t,s)=(t-m_K^2)\,\mathcal{P}_{Regge}^K$, which corresponds with the Reggeization of the contact term. In Eq. (\ref{eq:Mc}), $m_c^\mu$ is generally a  Kroll-Ruderman-type bare contact current which results from an elementary four-point Lagrangian, $G$ is an operator describing the coupling structure of the hadron vertex, and the auxiliary current $C^\mu$ is given by Eq. (A2) in Ref.~\cite{HH}.
The Regge current is required to fulfil the generalized Ward-Takahashi identity
\begin{eqnarray}
k_\mu \mathbb{M}^\mu &=& \Delta_K^{-1}(p_K)e\Delta_K(p_K-k)F_{R,t} \\
&-& F_s\,S_p(p+k)eS_p^{-1}(p) - e F_{R,t} + e F_s + k_\mu \mathbb{M}^\mu_{int}\,,\nonumber
\end{eqnarray}
which warrants its gauge invariance if the contact current is constructed to have the property
\begin{equation}
k_\mu \mathbb{M}^\mu_{int} = e (F_{R,t} - F_s).
\label{divergence}
\end{equation}
A specific form of the contact current depends on the chosen coupling in the strong vertex. Here we consider two possible forms of this coupling.

\subsubsection{Pseudoscalar coupling in the strong vertex}
Assuming the pseudoscalar (PS) coupling in the 
$K^+\!\Lambda p$ interaction vertex, $G_{PS}=g_{K\Lambda p}\gamma_5$, the gauge non-invariant term of the $K^+(494)$ exchange in the \emph{t} channel reads
\begin{equation}
- e g_{K\Lambda p} \gamma_5 \frac{f_t(t)}{t-m_K^2}(t-m_K^2)\frac{k\cdot \varepsilon}{k^2},
\label{eq:t}
\end{equation}
where $f_t(t)$ is a hadron form factor. After Reggeizing this contribution, when $f_t(t)/(t-m_K^2)$ turns into $\mathcal{F}_t(t,s)/(t-m_K^2)$, the term (\ref{eq:t}) has the form
\begin{equation}
- e g_{K\Lambda p} \gamma_5 \mathcal{F}_t(t,s)\frac{k\cdot \varepsilon}{k^2}.
\label{eq:t-Regge}
\end{equation}
In the \emph{s} channel, the gauge non-invariant term reads
\begin{equation}
e g_{K\Lambda p} \gamma_5 f_s(s) \frac{k\cdot \varepsilon}{k^2},
\label{eq:s}
\end{equation}
where $f_s(s)$ is a hadron form factor.  These two gauge-invariance violating terms are annihilated with the contact current (\ref{eq:Mc}).
The bare contact current $m_c^\mu$, in the case of hadron form factors being unity, must satisfy the condition
\begin{equation}
k_\mu m_c^\mu = e(F_t - F_s) = (e G_{PS} - G_{PS} e) = 0 
\label{barecontact}
\end{equation}
where $F_t$ and $F_s$ are strong vertex factors in the $t$ and $s$ channels, respectively. Since in the PS coupling the strong vertex factors in both channels coincide, \emph{i.e.} $F_t=F_s=G_{PS}$, the longitudinal part of the contact current is determined solely by the second term in Eq. (\ref{eq:Mc}). This term can be given by Eq. (A2) in Ref.~\cite{HH} for $e_m=e_b=e$ and $e_{b^\prime}=0$, and assuming only the $s$- and $t$-channel contributions, \emph{i.e.}  $\delta_s=\delta_t=1$ and $\delta_u=0$,
\begin{equation}
\begin{split}
& G_{PS} C^\mu\varepsilon_\mu  \\
& \!\!\! = e g_{K\Lambda p}\gamma_5 \Bigg\{ -(2p_K-k)^\mu \frac{\mathcal{F}_t-1}{t-m_K^2}f_s-(2p+k)^\mu\frac{f_s-1}{s-m_p^2}\mathcal{F}_t \\
& + \hat{A}\,(1-f_s) (1-\mathcal{F}_t)\left[ \frac{(2p_K-k)^\mu}{t-m_K^2} + \frac{(2p+k)^\mu}{s-m_p^2} \right] \Bigg\}\varepsilon_\mu \\
& \!\!\! = e g_{K\Lambda p} \gamma_5 \Bigg\{ -2 \frac{\mathcal{F}_t-1}{t-m_K^2}f_s (\mathcal{M}_2-\mathcal{M}_3) + (\mathcal{F}_t-1)f_s\frac{k\cdot \varepsilon}{k^2} \\
& - 2 \frac{f_s-1}{s-m_p^2}\mathcal{F}_t \,\mathcal{M}_2 - (f_s-1)\mathcal{F}_t\frac{k\cdot\varepsilon}{k^2}\\
& + 2\hat{A}\,(1-f_s)(1-\mathcal{F}_t)\left[ \frac{\mathcal{M}_2-\mathcal{M}_3}{t-m_K^2} + \frac{\mathcal{M}_2}{s-m_p^2} \right] \Bigg\},
\end{split}
\label{eq:gce}
\end{equation}
where the second expression comprises the explicitly gauge invariant structures ${\cal  M}_i$ defined in Ref.~\cite{SB16}. It is evident  that all the gauge-violating terms proportional to $k\cdot \varepsilon / k^2$ in formulas (\ref{eq:t-Regge}), (\ref{eq:s}) and (\ref{eq:gce}) cancel each other and that the four-divergence of the contact term  (\ref{eq:Mc}) is
\begin{equation}
 k_{\mu}\mathbb{M}_{int}^{\mu}=eg_{K\Lambda p}\gamma_5(\mathcal{F}_t-f_s),
 \label{intPS}
\end{equation}
which agrees with the requirement in Eq.~(\ref{divergence}) as in this case  $F_{R,t}=g_{K\Lambda p}\gamma_5\mathcal{F}_t$ and $F_s=g_{K\Lambda p}\gamma_5f_s$. Contributions of the contact term to the scalar amplitudes, see Ref.~\cite{SB16} for more details on the general formalism, can be found in Appendix A.

The function \textcolor{red}{$\hat{A}$} in (\ref{eq:gce}) is an arbitrary phenomenological function that is constrained only by a condition to vanish at high energies. This condition is necessary to prevent the ``violation of scaling behaviour"~\cite{Drell}. Our choice for \textcolor{red}{$\hat{A}\equiv\hat{A}(s)$} is a ``dipole" shape
\begin{equation}
\hat{A}(s) = A_0 \frac{\Lambda_c^4}{\Lambda_c^4 + (s-s_{thr})^2},
\label{eq:astu}
\end{equation}
where $s_{thr} = (m_\Lambda + m_K)^2$ and $A_0$ and $\Lambda_c$ are free parameters giving the strength of this term and cutting it off (thereby limiting the affected region), respectively.

\subsubsection{Pseudovector coupling in the strong vertex}
In the case of pseudovector (PV) coupling, the $K^+\!\Lambda p$ vertex reads
$$G_{PV} = - \frac{g_{K\Lambda p}}{m_\Lambda+m_p} K^\mu \gamma_\mu \gamma_5,$$
where the momentum $K$ corresponds to the kaon field coming out of the strong vertex. In the $s$ channel, it holds $K = p_K$, whereas in the $t$ channel it is $K = p-p_\Lambda$.

Gauge non-invariant terms coming from the electric part of the $s$-channel contribution with PV coupling read
\begin{equation}
e g^\prime_{K\Lambda p} \gamma_5 f_s(s)\,[ (m_\Lambda+m_p) + \not k ]\,\frac{k\cdot \varepsilon}{k^2}
\label{eq:s-pv-gni}
\end{equation}
and the gauge non-invariant terms from the Reggeized $t$-channel $K^+(494)$ exchange read
\begin{equation}
-e g^\prime_{K\Lambda p} \gamma_5\,\mathcal{F}_t(t,s)\,(m_\Lambda+m_p)\, \frac{k\cdot \varepsilon}{k^2},
\label{eq:t-pv-gni}
\end{equation}
with $g^\prime_{K\Lambda p} = g_{K\Lambda p}/(m_\Lambda+m_p)$. 
 
Without any $u$ channel contribution and without the form factors, the bare contact current $m_c^\mu$ must again satisfy the condition (7) in Ref.~\cite{HH} which now reads
\begin{eqnarray}
k_\mu m_c^\mu = eg^\prime_{K\Lambda p} (\not \! p \,\, - \not \! p_\Lambda \,- \not \! p_K) \gamma_5 = eg^\prime_{K\Lambda p} \not\! k \gamma_5.
\end{eqnarray} 
In contraction with the polarization vector $\varepsilon_\mu(k)$, its contribution to the amplitude can be recast as
\begin{equation}
\varepsilon_\mu m_c^\mu = -eg^\prime_{K\Lambda p} \gamma_5 \!\not\! k \frac{k\cdot \varepsilon}{k^2}\,.
\label{eq:mc-pv2}
\end{equation}
This form assures gauge invariance in the case of no hadron form factors introduced because it cancels the term proportional to  $\not\! k$ in the proton-exchange contribution, see Eq.~(\ref{eq:s-pv-gni}) with $f_s=1$. \textcolor{blue}{The contribution of the bare contact current with a hadron form factor and after Reggeization reads}
\begin{equation}
\varepsilon_\mu m_c^\mu \mathcal{F}_t(t,s) = -eg^\prime_{K\Lambda p} \gamma_5 \!\not\! k \frac{k\cdot \varepsilon}{k^2}\mathcal{F}_t(t,s),
\label{eq:mc-pv}
\end{equation}
which now cancels out with the same term, just with the opposite sign, in Eq. (\ref{eq:gce-pv}) below.
Contrary to the PS-coupling case, with the PV coupling the contact current is determined by both terms in Eq. (\ref{eq:Mc}) where the second one now has the form
\begin{widetext}
\begin{equation}
\begin{split}
G_{PV}C^\mu\varepsilon_\mu &= {}\,\, g^\prime_{K\Lambda p}\gamma_5 \not\! p_K \, C^\mu\varepsilon_\mu = g^\prime_{K\Lambda p}\gamma_5 (\not\! p \,+ \not\! k \,- \not\! p_\Lambda)C^\mu\varepsilon_\mu
=g^\prime_{K\Lambda p}\gamma_5 (m_\Lambda+m_p+\not \!k)C^\mu\varepsilon_\mu
= G_{PS}C^\mu\varepsilon_\mu\\ &+ g^\prime_{K\Lambda p}\gamma_5 \not\! k \, C^\mu\varepsilon_\mu 
\! = G_{PS}C^\mu\varepsilon_\mu + e g^\prime_{K\Lambda p}\gamma_5  \Bigg\{ -2 \frac{\mathcal{F}_t-1}{t-m_K^2}f_s \left( -\mathcal{M}_4 + \mathcal{M}_5 + \frac{1}{k^2}\left(k\cdot p_\Lambda - k\cdot p\right) \mathcal{M}_6\right) \\
& + (\mathcal{F}_t-1)f_s \not k \,\frac{k\cdot \varepsilon}{k^2} + 2 \frac{f_s-1}{s-m_p^2} \mathcal{F}_t \left( \mathcal{M}_4 +\frac{k\cdot p}{k^2} \mathcal{M}_6 \right) - (f_s-1)\mathcal{F}_t \not k\,\frac{k\cdot \varepsilon}{k^2} \\
& + 2 \hat{A}(1-f_s)(1-\mathcal{F}_t) \Bigg[ \frac{1}{t-m_K^2} \left( -\mathcal{M}_4+\mathcal{M}_5  + \frac{1}{k^2}\left(k\cdot p_\Lambda - k\cdot p\right) \mathcal{M}_6\right) - \frac{1}{s-m_p^2} \left( \mathcal{M}_4 + \frac{k\cdot p}{k^2} \mathcal{M}_6 \right)\Bigg]\Bigg\}.
\end{split}
\label{eq:gce-pv}
\end{equation}
\end{widetext}
It is evident again that all the gauge-violating terms proportional to $k\cdot \varepsilon / k^2$ in formulas (\ref{eq:s-pv-gni}),  (\ref{eq:t-pv-gni}), (\ref{eq:mc-pv}), and (\ref{eq:gce-pv}), get mutually canceled which guarantees gauge invariance of the full Regge current. Corresponding formulas for the scalar amplitudes can be found in Appendix A.

\subsubsection{Comparison of pseudoscalar and pseudovector couplings}
The newly constructed Regge model (\emph{i.e.} a background part of the photoproduction amplitude), therefore, consists of Reggeized $t$-channel contributions of $K^+(494)$ and $K^{*}(892)$ trajectories with no hadron form factor, $s$-channel proton exchange with a hadron form factor $f_s(s)$ and with a standard Feynman propagator, and a contact term with the Regge residual function $\mathcal{F}_t(t,s)$ and hadron form factor $f_s(s)$.

%
%
\begin{figure}
    \centering
    \includegraphics[width=0.52\textwidth]{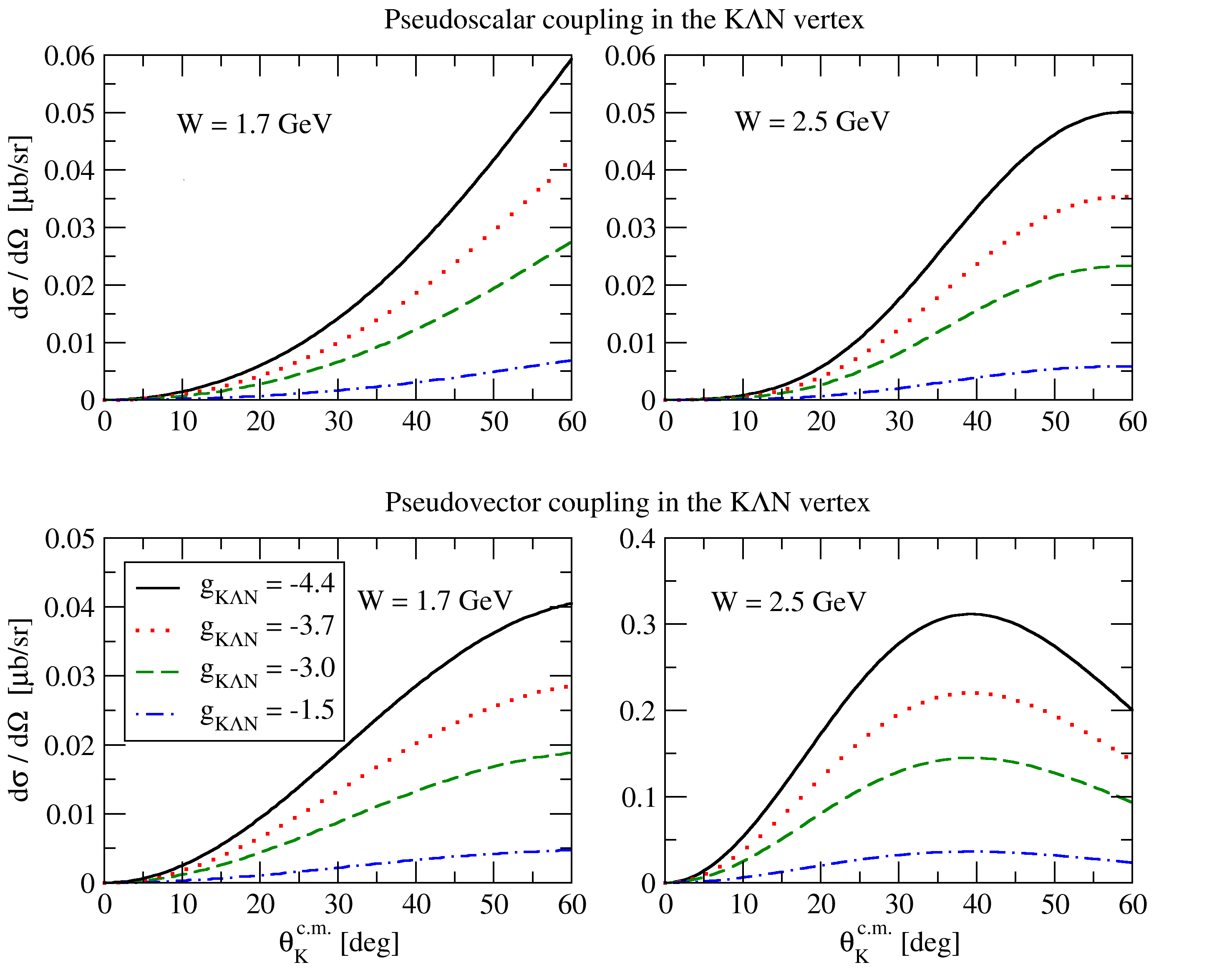}
    \caption{Cross section at forward kaon angles as given by the sole contact current without the transversal term in the case of pseudoscalar (upper row) and pseudovector (lower row) coupling in the $K\Lambda N$ vertex for various values of the governing coupling parameter $g_{K\Lambda N}$. Calculations done with the multidipole hadron form factor with $\Lambda_{bgr}=1.5\,\text{GeV}$ are shown for a near-threshold region ($W=1.7\,\text{GeV}$) and for a transitional region ($W=2.5\,\text{GeV}$).}
    \label{fig:PS-PV}
\end{figure}
%
%

\begin{figure*}
    \centering
    \includegraphics[width=0.75\textwidth]{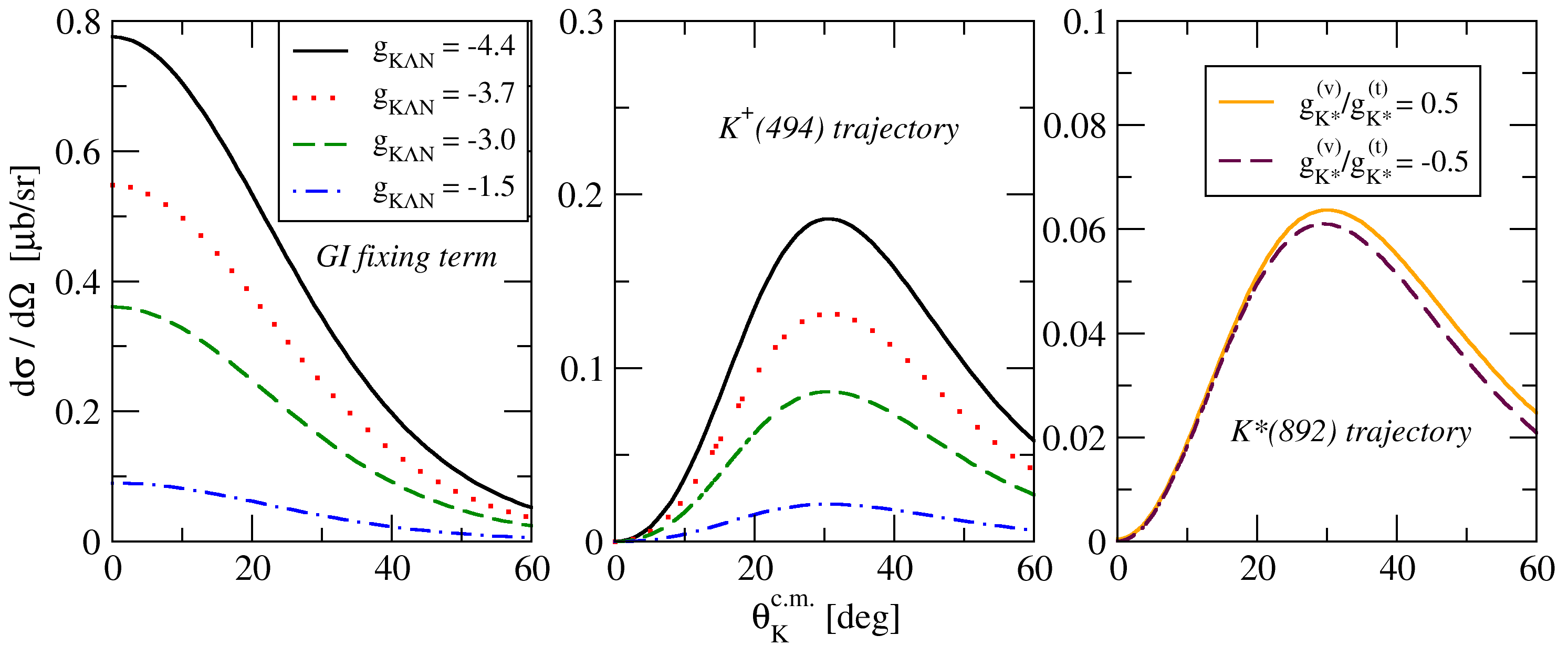}
    \caption{Cross section at forward kaon angles and for $W=2.5\,\text{GeV}$ as given solely by the gauge-invariance fixing term (Reggeized electric part of the $s$-channel proton exchange; left panel), $K^+(494)$ trajectory (center), and $K^*(892)$ trajectory (right) in the GLV method for gauge-invariance restoration.}
    \label{fig:KS-K-GI}
\end{figure*}

Before closing this section we deem important to sketch the difference the choice of either PS or PV coupling in the strong vertex makes. We do not observe any notable changes in the behaviour of either of the kaon trajectories. \textcolor{blue}{The proton exchange contributes with approximately the same magnitude in both types of coupling and only its shape varies mildly: With the PS coupling the proton exchange contribution decreases with increasing kaon angle, while with the PV coupling it increases with growing kaon angle. In any case, we do not observe as strong a dependence of the proton exchange on the $g_{K\Lambda N}$ value as it is revealed when we introduce the Guidal, Laget, and Vanderhaeghen (GLV) method for gauge-invariance restoration (see Fig.~\ref{fig:KS-K-GI} and discussion in the next Section).} What we do observe, however, is a transformation of the contact current contribution, see Fig.~\ref{fig:PS-PV}. With the PS coupling this contribution increases steadily, whereas with the PV coupling the contact current  \textcolor{blue}{contribution rises mildly with growing kaon angle  saturating at some point}. This effect is more pronounced at the higher end of the resonance region, $W=2.5\,\text{GeV}$, where contact currents with different type of coupling produce quite different  shapes in the cross section. Note also the difference in magnitude \textcolor{blue}{at higher energies} of the contributions with differing coupling the origin of which is the cut-off parameter value of $\Lambda_{bgr}=1.5\,\text{GeV}$ for the hadron form factor used in these calculations \textcolor{red}{(Eq.~(\ref{eq:Fd}) shows how this cut-off parameter enters the calculation of the dipole hadron form factor for the background if one interchanges $\Lambda_R$ with $\Lambda_{bgr}$ and inserts the proton mass instead of the resonance mass $m_R$)}. Whereas this value is the usual resulting value in fits with the PS coupling, a background with the PV coupling usually needs a smaller cut-off value to be accordingly suppressed.

\subsubsection{Gauge invariance restoration with the GLV method}
\textcolor{blue}{
A very popular recipe for the gauge-invariance restoration in the Regge model, which is used quite often in the literature, is the GLV method introduced in Ref.~\cite{Guidal}. In order to restore the gauge invariance, these authors replenish the Reggeized $t$-channel contributions of $K$ and $K^*$ trajectories with the electric part of the $s$-channel Born diagram which compensates the gauge-breaking term stemming from the $t$-channel diagram where the kaon is exchanged. Even though this method can provide relatively good data description, it was recently revealed that it cannot be obtained from field theory in any approximation~\cite{HH}. What is more, the presence of the gauge-restoring $s$-channel proton exchange seems to have a decisive effect on description of the differential cross section at very forward angles, as we show in Fig.~\ref{fig:KS-K-GI}. Contributions of both the $s$-channel term for fixing the gauge invariance (left panel in Fig.~\ref{fig:KS-K-GI}) and the $K^+(494)$ trajectory (center panel) are governed by the $g_{K\Lambda N}$ coupling constant. While the peak created by the kaon trajectory merely intensifies with decreasing value of the $g_{K\Lambda N}$, the gauge-invariance fixing term changes from a plateau-like behaviour to a steeply decreasing behaviour with kaon angle. In some cases, that strong a contribution of the gauge-invariance fixing term may result into a steep rise at very small kaon angles, which is however not observed in experiments (see \emph{e.g.} Fig.~5.11 in Ref.~\cite{DeCruz-PhD}). The $K^*(892)$ trajectory does not to seem to play an important role in changing the shape of the cross-section prediction at forward kaon angles, nor does it markedly change its behaviour when the relative sign of its couplings is switched.
}

\textcolor{blue}{
We stress that it is our strong need for a reliable description of the region of very forward kaon angles, being substantiated by our aim to use the RPR models further on for hypernuclei calculations, which forces us to reconsider the whole idea of gauge invariance restoration in the Regge model. As illustrated in Figs.~\ref{fig:PS-PV} and \ref{fig:KS-K-GI}, the choice of the correct method of restoring gauge invariance strongly affects the dynamics in this very kinematic area and, therefore, is of the utmost importance for revealing the physical mechanisms entering the game.
}

\section{Regge-plus-resonance Model}
\label{sec:RPR}
Although the Regge theory is a high-energy tool by construction, it can reproduce the order of magnitude of the forward-angle pion and kaon photoproduction \cite{RPR} and kaon electroproduction \cite{Guidal2} observables remarkably well even in the resonance region. Nevertheless, it is evident that a pure non resonant description, such as the Regge model, cannot be expected to describe the reaction at energies in the resonance region \cite{RPR}. The cross section near threshold exhibits structures, such as peaks at certain energies, which might reflect the presence of individual resonances. These are incorporated into the Regge-plus-resonance (RPR) model by extending the Reggeized background with $s$-channel diagrams with exchanges of nucleon resonances. For these diagrams, standard Feynman propagators are assumed where, as in the isobar approach, the resonance finite lifetime is taken into account through the substitution
$$s-m_R^2 \rightarrow s-m_R^2+\texttt{i}m_R\Gamma_R$$
in the propagator denominator with the $m_R$ and $\Gamma_R$ the mass and width of the propagating state, respectively. For more details on formalism for exchanges of nucleon resonances with spin up to 5/2 we refer to our work in Ref.~\cite{SB16}.

In order to retain the RPR approach reasonable, the resonance contributions should vanish in the high-energy region. This is achieved with the help of hadron form factors which should be strong enough not to allow the resonant terms to contribute beyond the resonance region. For this purpose, one usually opts for a multidipole, $F_{\text{md}}$, or multidipole-Gaussian, $F_{\text{mdG}}$, shape of the hadron form factor,
\begin{subequations}
\begin{align}
F_{\text{md}}(x,m_R,\Lambda_R,J_R) = & F_{\text{d}}^{J_R+1/2}(x,m_R,\Lambda_R),\label{eq:Fmd}\\
\begin{split}F_{\text{mdG}}(x,m_R,\Lambda_R,J_R,\Gamma_R) = 
& F_{\text{d}}^{J_{R}-1/2}(x,m_R,m_R\tilde{\Gamma}_R) \\
& \times F_{\text{G}} (x,m_R,\Lambda_R),\label{eq:FmG}\end{split}
\end{align}
\label{eq:FFmd-mdG}
\end{subequations}
where $m_R$, $\Gamma_R$, $J_{R}$, $\Lambda_R$, and $x\equiv s,t,u$ stand for the mass, width, and spin of the particular resonance, cut-off parameter of the form factor, and Mandelstam variables, respectively, and $\tilde{\Gamma}_R$ is a modified decay width,
\begin{equation}
\tilde{\Gamma}_R(J_R) = \frac{\Gamma_R}{\sqrt{2^{1/2J_R}-1}}.
\end{equation}
These two form factors fall off with energy much more sharply than the dipole, $F_{\text{d}}$, or the Gaussian, $F_{\text{G}}$, form factors,
\begin{subequations}
\begin{align}
F_{\text{d}} (x,m_R,\Lambda_R) = & \frac{\Lambda_R^4}{(x-m_R^2)^2+\Lambda_R^4},\label{eq:Fd}\\
F_{\text{G}} (x,m_R,\Lambda_R) = & \exp [-(x-m_R^2)^2/\Lambda_R^4].\label{eq:FG}
\end{align}
\label{eq:FFd-G}
\end{subequations}
With help of the multidipole or the multidipole-Gaussian form factors, only the Regge part of the amplitude remains in the high-energy region. \textcolor{red}{The hadron form factors introduced into the kaon and proton exchanges, $f_t$ and $f_s$  in Eqs.~(\ref{eq:t}) and (\ref{eq:s}), respectively, are also expressed as $f_s,\,f_t\,=\,F_{\text{x}}$ where $\text{x\,=\,md, mdG, d,}$ and G. Note that after Reggeization only $f_s$ remains in the amplitude as $f_t$ was replaced with the Regge residual function ${\cal F}_t$.
}

\begin{table}[h!]
\caption{Meson and baryon resonances which are included in the description of the $p(\gamma, K^+)\Lambda$ process. For each resonance, the mass, width, spin, parity, and status are shown. Masses and widths are precisely the values which we use in the present work and are in concert with values presented in the Particle Data Tables 2018~\cite{PDG}.} 
\hspace{.0cm}
\begin{tabular*}{\columnwidth}{l @{\extracolsep{\fill}} c c c c c r}
\hline\hline
&                   &                & Mass        & Width       &           & \\
& Nickname          & Particle       & [MeV]       & [MeV]       & $J^\pi$   & Status\\  \hline
& N3\hfill\vadjust{}& $S_{11}(1535)$ & 1520        & 110         & $1/2^-$   & ****\\
& N4\hfill\vadjust{}& $S_{11}(1650)$ & 1670        & 100         & $1/2^-$   & ****\\
& N5\hfill\vadjust{}& $D_{13}(1700)$ & 1750        & 150         & $3/2^-$   & ***\\
& N7\hfill\vadjust{}& $P_{13}(1720)$ & 1690        & 325         & $3/2^+$   & ****\\
& N8\hfill\vadjust{}& $D_{15}(1675)$ & 1675        & 150         & $5/2^-$   & ****\\
& N9\hfill\vadjust{}& $F_{15}(1680)$ & 1675        & 130         & $5/2^+$   & ****\\
& P1\hfill\vadjust{}& $P_{11}(1880)$ & 1870        & 235         & $1/2^+$   & ***\\
& P2\hfill\vadjust{}& $P_{13}(1900)$ & 1900        & 400         & $3/2^+$   & ****\\
& P3\hfill\vadjust{}& $F_{15}(2000)$ & 2000        & 380         & $5/2^+$   & **\\
& P4\hfill\vadjust{}& $D_{13}(1875)$ & 1900        & 220         & $3/2^-$   & ***\\
& P5\hfill\vadjust{}& $F_{15}(1860)$ & 1860        & 170         & $5/2^+$   & **\\
& M3\hfill\vadjust{}& $D_{15}(2570)$ & 2570        & 250         & $5/2^-$   & **\\
& M4\hfill\vadjust{}& $D_{15}(2060)$ & 2130        & 350         & $5/2^-$   & ***\\
\hline\hline
\end{tabular*}
\label{TableAllRes} 
\end{table}
\begin{center}
\begin{table*}
\caption{An overview of $N^*$'s included in various recent models. Our current RPR-BS models are compared with our older RPR-1 and RPR-2 models~\cite{RPR-12}, with the Ghent RPR-2011 model and also with our recent isobar models BS1~\cite{SB16}, BS2~\cite{SB16}, and BS3~\cite{SB18}.}
\begin{center}
\begin{tabular*}{\textwidth}{l @{\extracolsep{\fill}} c c c c c c c c c c c c c c c}
\hline \hline
           &  N3 & N4 & N5 & N6 & N7 & N8 & N9 & P1 & P2 & P3 & P4 & P5 & M1 & M3 & M4 \\ \hline
RPR-BS     &  $\checkmark$ & $\checkmark$ & $\checkmark$ & & $\checkmark$ & $\checkmark$ & $\checkmark$  & & $\checkmark$ & $\checkmark$ & $\checkmark$ & $\checkmark$ & & $\checkmark$ & \\
RPR-BS(pv) &  $\checkmark$ & $\checkmark$ & $\checkmark$ & & $\checkmark$ & & $\checkmark$ & $\checkmark$ & $\checkmark$ & $\checkmark$ & $\checkmark$ & $\checkmark$ & & & $\checkmark$ \\
RPR-1 \& 2 &  $\checkmark$ & $\checkmark$ & & $\checkmark$ & $\checkmark$ & & & & & & $\checkmark$ & & & & \\
RPR-2011   &  $\checkmark$ & $\checkmark$ & & & $\checkmark$ & & $\checkmark$ & $\checkmark$ & $\checkmark$ & $\checkmark$ & $\checkmark$ & & & & \\
BS1        &  $\checkmark$ & $\checkmark$ & & & $\checkmark$ & & $\checkmark$ & & $\checkmark$ & $\checkmark$ & $\checkmark$ & $\checkmark$ & & & \\
BS2        &  $\checkmark$ & $\checkmark$ & & $\checkmark$ & $\checkmark$ & & $\checkmark$ & & $\checkmark$ & $\checkmark$ & $\checkmark$ & $\checkmark$ & & & \\
BS3        &  $\checkmark$ & $\checkmark$ & & $\checkmark$ & $\checkmark$ & & $\checkmark$ & & $\checkmark$ & $\checkmark$ & $\checkmark$ & $\checkmark$ & $\checkmark$ & & \\ \hline \hline
\end{tabular*}
\end{center}
\label{tab:N}
\end{table*}
\end{center}

\subsection{Nucleon resonances in the $s$ channel}
In order to select a set of nucleon resonances which describes the $p(\gamma,K^+)\Lambda$ data best, one has to carry out a great amount of fits assuming many combinations of $N^*$'s. First, we constructed a ``maximal" model including all nucleon resonances with spin up to 5/2 that might contribute in the $K^+\Lambda$ photoproduction \textcolor{blue}{(we did not include $N^*$'s with spin higher than 5/2 even though some authors recently claim, based on the multichannel partial-wave analysis, that a spin-7/2 state  $F_{17}(2200)$~\cite{Hunt} or $F_{17}(2300)$ in Ref.~\cite{Hunt2} might have a strong coupling to $K^+\Lambda$)}. That resulted into model with 14 $N^*$'s. After that we systematically omitted nucleon resonances one by one and checked the $\chi^2$ values and correspondence with data. During this process, we have also been slightly manually modifying the particles' masses and widths inside the PDG limits~\cite{PDG} (if given). From the values of particles' masses and widths shown in Table I in Ref.~\cite{SB16} we arrived at values only mildly different, which nonetheless play some role in reducing the $\chi^2$ value. Particularly, the masses of N3, N4, N5, N9, and P4 were shifted to values of 1520, 1670, 1750, 1675, and 1900~MeV, respectively. The widths of N3, N4, N7, P2, P3, and P5 were changed to respective values of 110, 100, 325, 400, 380, and 170~MeV. Note that the width of P2, whose value is much higher than the upper limit imposed by the PDG~\cite{PDG}, is inspired by a previous thorough analysis at Ghent University~\cite{DeCruz}. What is more, we introduced \textcolor{blue}{two new resonant states in the $D_{15}$ partial wave, N(2060)$5/2^-$ and N(2570)$5/2^-$}, which were not considered before, with masses \textcolor{blue}{2130 and 2570~MeV and widths 350 and 250~MeV, respectively. The $D_{15}(2570)$ state was observed in a recent multipoles analysis~\cite{MartSakinah} where it plays a rather important role. Moreover, nucleon resonances above 2200~MeV have been recently proposed in a partial-wave analysis using a multichannel framework~\cite{HuntPhD}. For a notation we use here we refer our reader to the Table~\ref{TableAllRes} which also summarizes the masses and widths used in the RPR-BS model. In the RPR-BS(pv) model, the masses and widths have the same values, except for $S_{11}(1535)$ and $F_{15}(1680)$, whose masses are slightly changed to 1510 and 1665~MeV, respectively.}

As we limit our study to the $K^+\Lambda$ channel only, we do not introduce any $\Delta$ resonances since their decay to this channel is prohibited by isospin conservation.

\section{Seeking the best fit}
\label{sec:fit}
As the Regge and Regge-plus-resonance frameworks are effective ones with coupling parameters and cutoff values of hadron form factors not determined, our primary goal is to adjust these parameters to experimental data. The parameters in need of adjustment are the coupling constants of the Regge background, \emph{i.e.} $g_{K\Lambda N}$, governing the behaviour of the $K^+(494)$ trajectory, contact current, and proton exhange in the $s$ channel, and the vector and tensor coupling constant $G_{K^{*}}^{(v,t)}$ of the $K^{*}(892)$ trajectory, and coupling constants of additional nucleon resonances. Each spin-1/2 resonance adds one free parameter, while higher-spin resonances add two free parameters. We also need to determine values of cutoff parameters for hadron form factors suppressing nucleon resonances (a common cutoff parameter $\Lambda_N$) and also the proton exchange (cutoff parameter $\Lambda_{bgr}$). Please note that the hadron form factor for the $s$-channel Born proton exchange is a novel feature of our model. Moreover, when we assume also the transversal term in the contact current, this gives us two more parameters, \emph{i.e} $A_0$ and $\Lambda_c$ in Eq. (\ref{eq:astu}). In total, we need to fixate around 25 free parameters depending mainly on the number of $N^*$'s we put in.

We adjusted the free parameters with help of the least-squares fitting procedure using the MINUIT code~\cite{Minuit}. Since it is well known that MINUIT uses a nonlinear transformation for the parameters with limits, making the accuracy of the resulting parameter worse as it approaches its limiting value, we introduced limiting values only to the background cutoff parameter $\Lambda_{bgr}$ of hadron form factor and to the $g_{K\Lambda N}$ coupling parameter. The latter one was further removed and the $g_{K\Lambda N}$ coupling was therefore allowed to violate the SU(3) symmetry slightly more than what is normally considered, i.e. within 20\% around the central value~\cite{SB16}. \textcolor{blue}{This more liberal approach of ours is motivated by the fact that the $K\Lambda N$ coupling effectively accounts for the coupling of the whole trajectory, {\it i.e.} also for higher-lying poles such as $K_1(1270)$. What is more, it seems that the $p(\gamma,K^+)\Sigma^0$ and $p(\gamma,K^0)\Sigma^+$ reactions can be well fitted only when $g_{K\Sigma N}$ coupling is allowed to vary far off the SU(3) limits and thus these photoproduction reactions might become an important source of information on the validity of the SU(3) relation~\cite{Kamano}.}

First, we have been fitting on high-energy data and adjusting only the Regge-background parameters $g_{K\Lambda N}$, $G_{K^*}^{(v)}$, and $G_{K^*}^{(t)}$, which resulted in finding one deep minimum. We deem that this hints to a strong probability for this minimum to be a global minimum. With high-energy data we mean cross-section data from the CLAS 2010 collaboration~\cite{CLAS10} for $W > 2.36\,\text{GeV}$ (230 data points), from the CLAS 2005 collaboration~\cite{CLAS05} for $E_\gamma^{lab}\geq 2.505\,\text{GeV}$ (95 data points), 20 recent data points from the LEPS collaboration~\cite{LEPS18} for $E_\gamma^{lab} \geq 2.55\,\text{GeV}$, \textcolor{blue}{and 305 hyperon polarization data points from the CLAS 2010 collaboration~\cite{CLAS10} for $W>2.23\,\text{GeV}$}. The data were limited not only to high energies but also to forward kaon angles only, \emph{i.e.} $\theta_K^{c.m.}\leq 60^\circ$, since this is the kinematical region where the $t$-channel Reggeistics takes place. Subsequently, we added data in the resonance region, namely the cross-section data from the CLAS 2010 collaboration~\cite{CLAS10} for $W<2.36\,\text{GeV}$ (1247 data points), the CLAS 2005 collaboration~\cite{CLAS05} for $E_\gamma^{lab}<2.5\,\text{GeV}$ (1037 data points), the LEPS collaborations for $E_\gamma^{lab}<2.38\,\text{GeV}$ (54 data points)~\cite{LEPS06} \textcolor{blue}{and $E_\gamma^{lab}<2.55\,\text{GeV}$ (40 data points)~\cite{LEPS18}}, and 91 cross-section data points collected by Adelseck and Saghai in their paper~\cite{AS} from various experimental facilities, hyperon-polarization data from the CLAS 2010 collaboration~\cite{CLAS10} for $W<2.23\,\text{GeV}$ (925 data points), \textcolor{blue}{and 314 hyperon-polarization data points from CLAS 2016~\cite{CLAS16}}. These data for adjusting $N^*$ parameters were naturally not restricted with respect to kaon angle $\theta_K^{c.m.}$. No weight factor was introduced to any data so they all come to the fitting process with the same importance. \textcolor{blue}{In total, we exploited 4358 experimental data points in our fitting procedure.} With this data set, we fitted the $N^*$'s coupling constants while keeping the background parameters on their values from the high-energy fit. However, soon we realized that we can achieve significantly better results when we fit all parameters simultaneously and we, therefore, merged all of these data sets into a single data file which we subsequently used for the rest of the fitting procedure.

An astute reader may have noticed that we did not include all differential-cross-section data available to us in these days. This is because there exists some ambiguities and inconsistencies between some of them, the most notable being the inconsistency between the extensive CLAS data set and the results from the SAPHIR~\cite{SAPHIR} collaboration. As pointed out and discussed in Ref.~\cite{Sarantsev}, a common fit to both data sets would be possible only after inclusion of a normalization function or factor. We do not consider this issue here (more details can be also found in Ref.~\cite{ByMa07}) and therefore restrict ourselves only to the CLAS data.

Another inconsistency apparently exists between CLAS data and data Boyarski \emph{et al.}~\cite{SLAC} from the Stanford Linear Accelerator Center (SLAC) which agree well in shape but the CLAS cross-section data are systematically lower than SLAC in scale by roughly a factor of two (even though a direct comparison is difficult since kinematics of both sets do not overlap much)~\cite{Dey}. As we know e.g. from the analysis made by Guidal \emph{et al.}~\cite{Guidal}, a model resulting from fitting SLAC high-energy data and projecting down to CLAS energies consistently overpredicts the $K^+ \Lambda$ cross sections (for illustration see Fig. 20 in Ref.~\cite{CLAS05}). Adjusting model parameters to the high-energy SLAC data and subsequently extrapolating them to the resonance region can, therefore, lead to a dissatisfactory description of resonance region (or it can strongly influence the interference pattern among background and resonant terms). Thus, we have decided not to use the SLAC data in our analysis.

Last, from the polarization data of the CLAS 2016~\cite{CLAS16} experiment, comprising around 1500 data on hyperon polarization $P$, target asymmetry $T$, beam asymmetry $\Sigma$, and double-polarization observables $O_x$ and $O_z$, \textcolor{blue}{we have included only the hyperon-polarization data to the data base for the fits. On the one hand, we realized that these data do not bring us closer to understanding the underlying mechanism of $N^*$'s interferences and their inclusion leads to higher values of $\chi^2$ in our fits since our models are not able to capture the very minute shapes these data present. On the other hand, the role of the CLAS 2016 hyperon-polarization data may be in giving more weight to the other hyperon-polarization data in our fits as they are both mutually consistent}. We did not include the rest of the CLAS 2016 data set as we reckon fitting to these data would be beyond the scope of the present work, whose main aim is to present a novel way to achieve the gauge invariance restoration. Nonetheless, we show these data and compare our model predictions with them in the next section in relevant figures.

For a much more thorough discussion of the fitting procedure, see Ref.~\cite{SB16}.

%
%
\begin{figure}[t]
\centering
\includegraphics[width=0.50\textwidth]{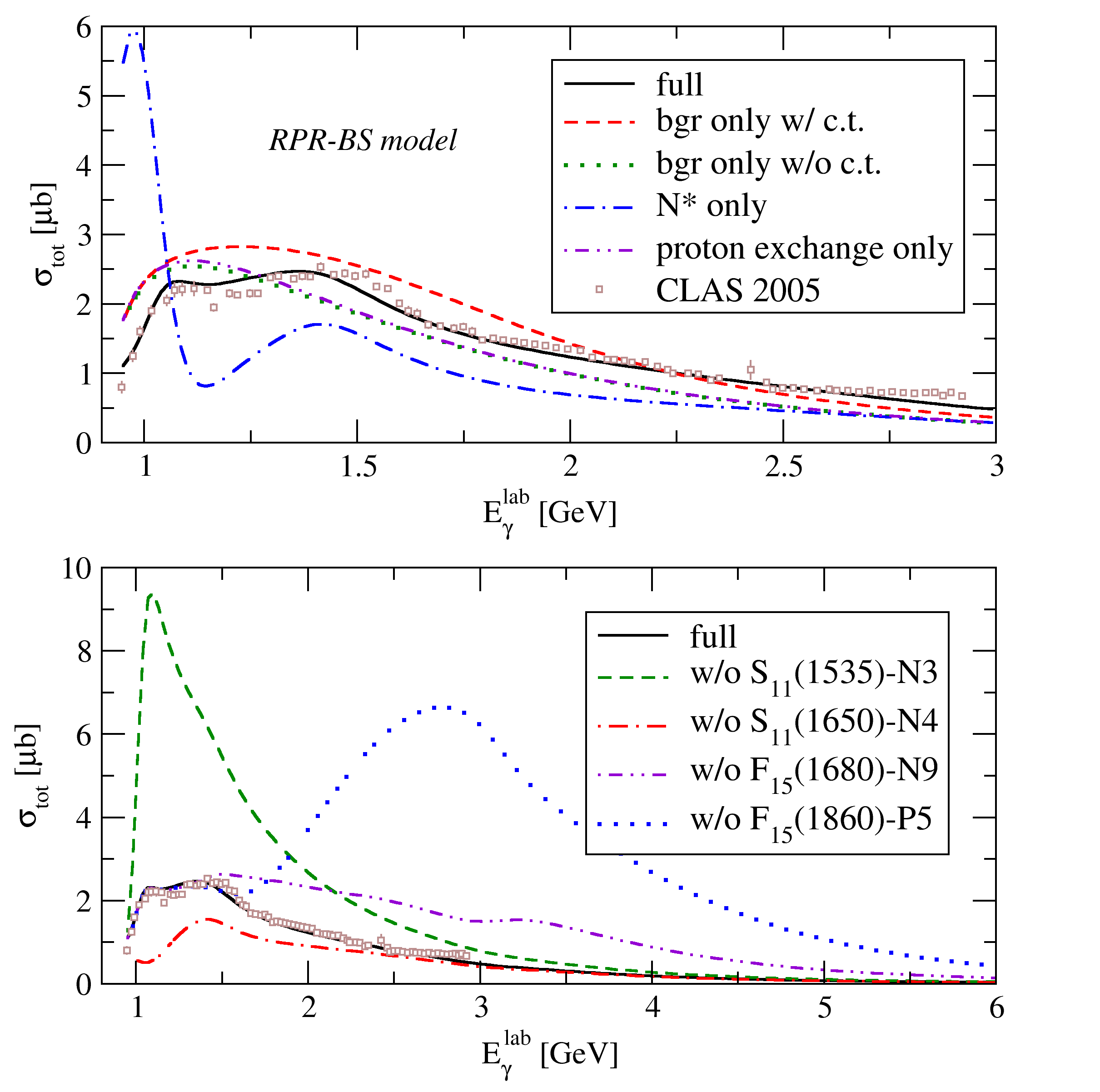}
\caption{Total cross-section prediction of the RPR-BS model (solid line) in comparison with CLAS 2005 data~\cite{CLAS05}. Contributions to the RPR-BS model from background (dashed line), background without the contact term (dotted line), and from all $N^*$'s (dash-dotted line) are illustrated in the upper figure. Behaviour of the RPR-BS model when a particular $N^*$ state is omitted is shown in the lower figure.}
\label{fig:totcrs}
\end{figure}

%
%
\begin{figure*}[ht]
    \centering
    \includegraphics[width=0.75\textwidth]{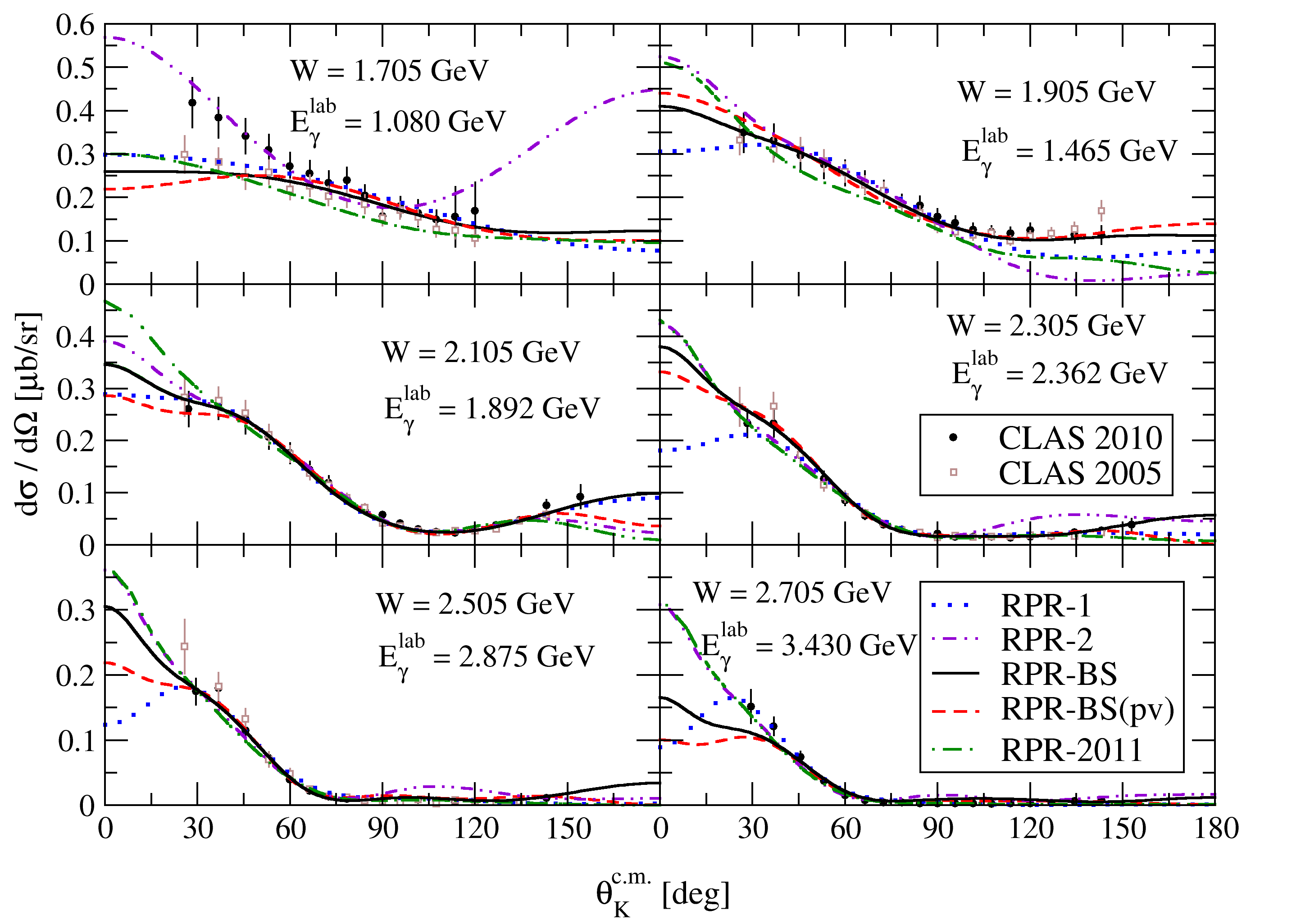}
    \caption{Angular dependence of the cross section calculated with RPR-BS (solid line), RPR-BS with pseudovector coupling in the $K\Lambda N$ vertex (dashed line), RPR-1~\cite{RPR-12} (dotted line), RPR-2~\cite{RPR-12} (dashed double-dotted line), and RPR-2011~\cite{DeCruz} (dash-dotted line) are shown for six values of the center-of-mass energy. The data stem from the CLAS 2005~\cite{CLAS05} and CLAS 2010~\cite{CLAS10} collaborations. }
    \label{fig:crs-th0}
\end{figure*}

\section{Discussion of results}
\label{sec:dis}
In this section, we present our new Regge-plus-resonance models for photoproduction of $K^+\Lambda$ and compare their predictions of the cross section, hyperon polarization, and two other double polarization observables with experimental data and results of other models.

The set of nucleon resonances in the current models \textcolor{blue}{with the pseudoscalar and pseudovector coupling coined RPR-BS and RPR-BS(pv), respectively, which relates to the type of model and authors' names,} is in a good concert with the set chosen in the Ghent RPR-2011 model~\cite{DeCruz} and also with $N^*$'s in our isobar models~\cite{SB16,SB18}, see the overview in Table~\ref{tab:N}. \textcolor{blue}{We also confirm conclusions from the recent multichannel partial-wave analysis by Hunt and Manley~\cite{Hunt2} that the states $S_{11}(1650)$, $P_{13}(1720)$, $P_{11}(1880)$, and $P_{13}(1900)$ contribute significantly to the $K^+\Lambda$ channel.} 
We partly corroborate the claim of Ref.~\cite{DeCruz} where authors found a decisive evidence for inclusion of $P_{11}(1880)$, $P_{13}(1900)$, and $F_{15}(2000)$ states as well as a compelling evidence for omitting $D_{13}(1700)$, $P_{11}(1710)$, and $D_{15}(1675)$ states as we include both $P_{13}(1900)$ and $F_{15}(2000)$ states and the $P_{11}(1880)$ state is included in the RPR-BS(pv) model. On the other hand, we do not include $P_{11}(1710)$ in both models and $D_{15}(1675)$ only in the RPR-BS(pv) model while including the $D_{13}(1700)$ state in both models. From the states which were not found to be important in the Ghent analysis, we include the \textcolor{blue}{$D_{15}(2570)$ and $D_{15}(2060)$ states (denoted as M3 and M4, respectively)}, even though their couplings are \textcolor{blue}{several} orders of magnitude smaller than couplings of other spin-5/2 resonances, and $F_{15}(1860)$. Both these resonant states play important roles at energies $W>2\,\text{GeV}$. \textcolor{blue}{We also feel the need for introducing two more states near the threshold, the $D_{13}(1700)$ and $D_{15}(1675)$ states, which were not found in the Ghent analysis to be crucial for data description.} 
The set of $N^*$'s in the RPR-BS(pv)  model is the same as the set in RPR-BS with only two exceptions: interchanging the \textcolor{blue}{$D_{15}(1675)$ and $D_{15}(2570)$} states for $P_{11}(1880)$ and $D_{15}(2060)$ states. In both of these models, we assume fixed decay widths of nucleon resonances. It is because when we introduce energy-dependent widths (as defined in Ref.~\cite{SB18}) the resulting cross-section prediction at small kaon angles and for $E_\gamma^{lab}>2\,\text{GeV}$ plummets drastically. The resonance width increases with increasing energy and the resonance thus moves away from the physical plane and contributes less (even though we detect a significant growth of coupling constants of some $N^*$'s in fits with energy-dependent widths, this seemingly cannot make the resonance influential enough); background terms are not strong enough at energies above 2~GeV and the overall result is then values of predicted cross section around 0.1~$\mu$b/sr, which is far less than what is observed and what other models produce.

When we compare the couplings of nucleon resonances in the RPR-BS model with their values in other models of ours (including the isobar BS models presented in Tab. \ref{tab:par}), we can see that couplings of $S_{11}(1535)$ and $S_{11}(1650)$ are in the RPR-BS model approximately twice as large as in other models while retaining the same sign, \textcolor{blue}{whereas in the RPR-BS(pv) model the $S_{11}(1535)$ changes its sign and the couplings of the $S_{11}(1650)$ have similar values as in the BS models}. \textcolor{blue}{Interestingly, the couplings of the $D_{13}(1875)$ state are in our RPR models an order of magnitude smaller in comparison with BS models and in the RPR-BS(pv) model its $G_1$ coupling even changes sign. Moreover, couplings of the $P_{13}(1720)$ state are in the BS1 and BS2 models opposite in comparison with the remainder of our models. } The rest of the coupling parameters show only minor changes. All parameters of our new models are summarized in Table~\ref{tab:par}.

\begin{table}[h!]
\caption{Coupling constants, parameters of the function $\hat{A}(s)$ in the transversal contact current, and cutoff values of hadron form factors of the final models are displayed. The cutoff values are shown in units of GeV. Errors of the parameters are included as well.}
\begin{center}
\begin{tabular*}{\columnwidth}{l @{\extracolsep{\fill}} d{1.6} d{1.6} d{1.4} d{1.4}}
\hline \hline
                              &\multicolumn{2}{c}{RPR-BS}        &\multicolumn{2}{c}{RPR-BS(pv)}        \\
		                      &\multicolumn{1}{l}{Value} &\multicolumn{1}{l}{Error} & \multicolumn{1}{c}{Value}           &\multicolumn{1}{l}{Error} \\ \hline
 $g_{K\Lambda N}/\sqrt{4\pi}$ & -2.251           & 0.029  & -2.124            & 0.032  \\
 $G_{K^*}^{(v)}$              & 0.023            & 0.003  &  0.014            & 0.003  \\
 $G_{K^*}^{(t)}$              & -0.049           & 0.007  & -0.029            & 0.006  \\
 $A_0$                        & -2.717           & 0.070  & -0.529            & 0.003  \\
 $\Lambda_c$                  & 1.203            & 0.007  &  1.429            & 0.013  \\
 $\Lambda_{bgr}$              & 1.958            & 0.009  &  1.235            & 0.024  \\
 $\Lambda_{N}$                & 1.966            & 0.009  &  1.864            & 0.015  \\ 
 $G(N3)$                      & 0.435            & 0.005  & -0.305            & 0.006  \\
 $G(N4)$                      & -0.144           & 0.001  & -0.038            & 0.001  \\
 $G_1(N5)$                    & 0.139            & 0.005  &  0.140            & 0.006  \\
 $G_2(N5)$                    & 0.009            & 0.004  & -0.006            & 0.005  \\
 $G_1(N7)$                    & 0.025            & 0.002  &  0.092            & 0.003  \\
 $G_2(N7)$                    & 0.039            & 0.001  &  0.051            & 0.001  \\ 
 $G_1(N8)$                    & 0.002            & 0.0001 &  -                & -      \\
 $G_2(N8)$                    & -0.007           & 0.0002 &  -                & -      \\
 $G_1(N9)$                    & -0.041           & 0.002  &  0.013            & 0.001  \\
 $G_2(N9)$                    & 0.014            & 0.002  & -0.058            & 0.001  \\
 $G(P1)$                      & -                & -      & -0.229            & 0.006  \\
 $G_1(P2)$                    & 0.015            & 0.001  &  0.017            & 0.002  \\
 $G_2(P2)$                    & -0.025           & 0.0004 & -0.019            & 0.001  \\
 $G_1(P3)$                    & -0.023           & 0.0002 & -0.016            & 0.0002 \\
 $G_2(P3)$                    & 0.019            & 0.0002 &  0.012            & 0.0002 \\
 $G_1(P4)$                    & 0.063            & 0.004  & -0.042            & 0.003  \\
 $G_2(P4)$                    & 0.109            & 0.0003 &  0.037            & 0.003  \\
 $G_1(P5)$                    & 0.055            & 0.001  &  0.018            & 0.001  \\
 $G_2(P5)$                    & -0.036           & 0.001  &  0.010            & 0.001  \\
 $G_1(M3)$                    & -0.00005         & 0.00003 &  -               & -      \\
 $G_2(M3)$                    & -0.000008        & 0.00004 &  -               & -      \\
 $G_1(M4)$                    & -                & -      &  0.0005           & 0.0002 \\
 $G_1(M4)$                    & -                & -      & -0.001            & 0.0004 \\
 $\chi^2/\text{n.d.f.}$       & 1.69             & -      &  1.74             & -      \\ \hline \hline 

\end{tabular*}
\end{center}
\label{tab:par}
\end{table}
%
%
\begin{figure*}
    \centering
    \includegraphics[width=0.75\textwidth]{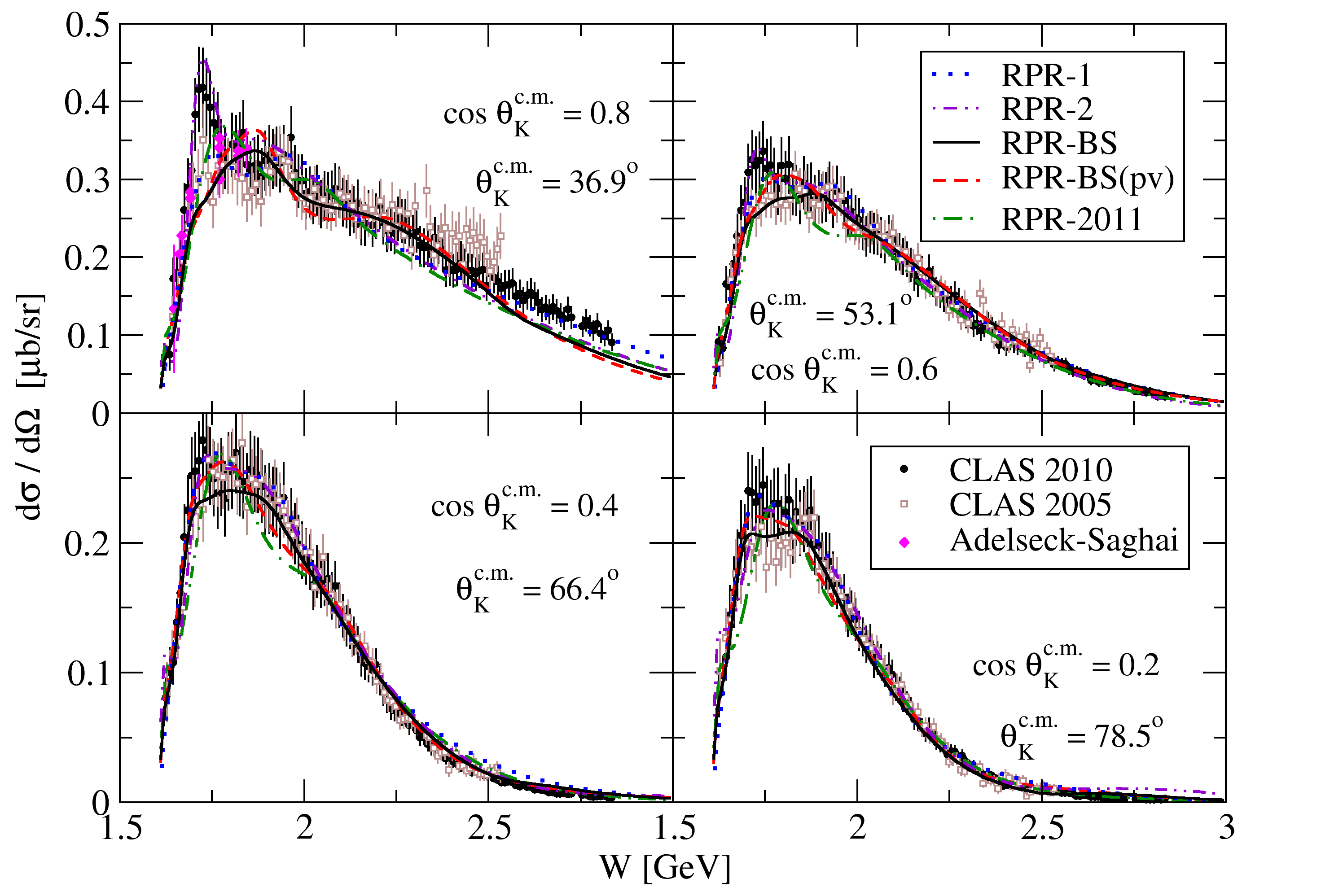}
    \caption{Results for the energetic dependence of the cross section are shown for four various kaon angles. The data are from the CLAS 2005~\cite{CLAS05}, CLAS 2010~\cite{CLAS10}, LEPS 2006~\cite{LEPS06} and LEPS 2018~\cite{LEPS18} collaborations and from the collection of Ref.~\cite{AS}. }
    \label{fig:crs-w}
\end{figure*}
%
%
\begin{figure}[b]
    \centering
    \includegraphics[width=0.51\textwidth]{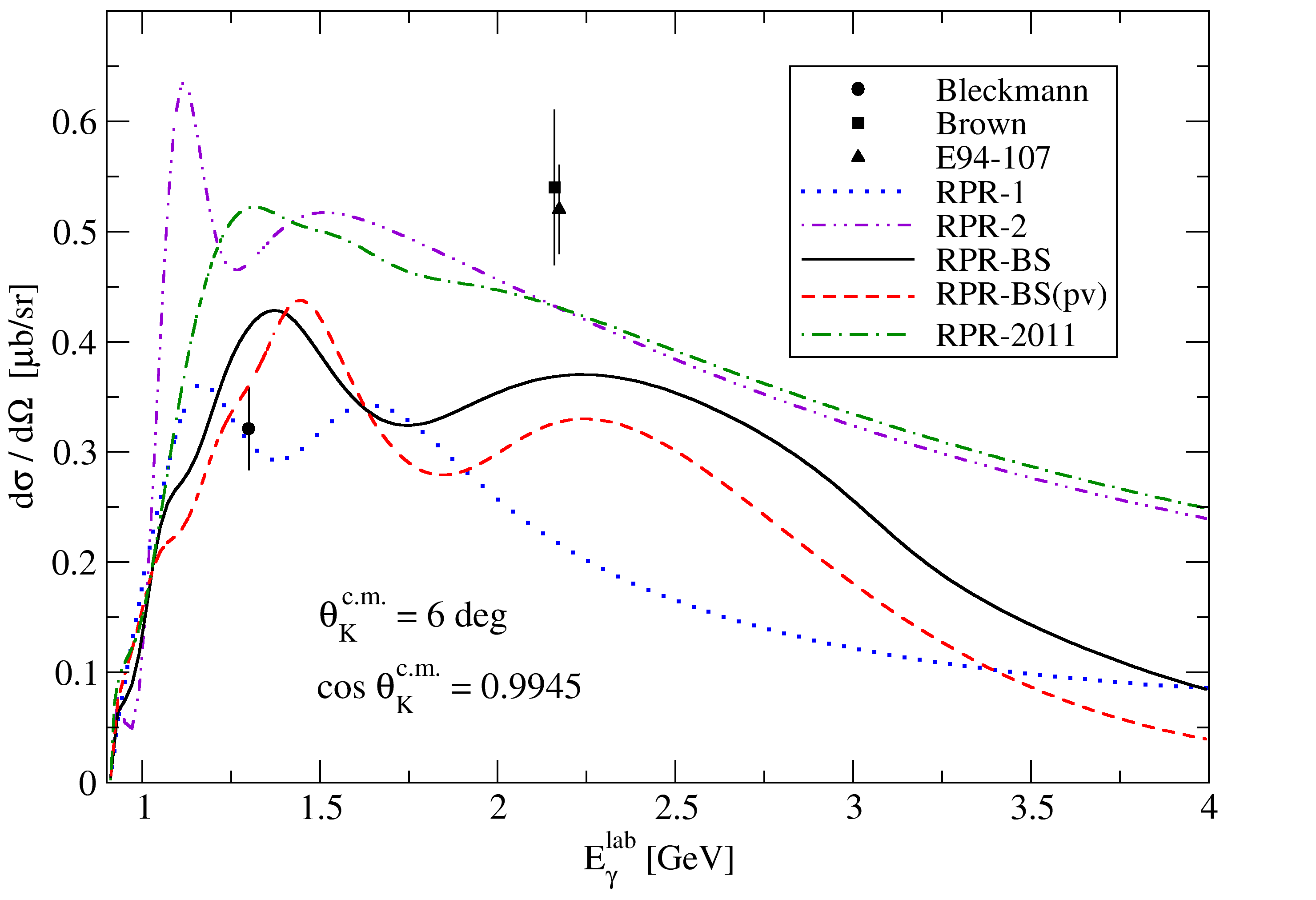}
    \caption{Differential cross section of $p(\gamma,K^+)\Lambda$ for $\theta_K^{c.m.}=6^\circ$ is shown. Predictions of the models are compared with data of Bleckmann \emph{et al.}~\cite{Bleck}, Brown \emph{et al.}~\cite{Brown}, and experiment E94-107~\cite{E94-107}. }
    \label{fig:crs-6deg}
\end{figure}

%
%

\begin{figure}[b]
    \centering
    \includegraphics[width=0.51\textwidth]{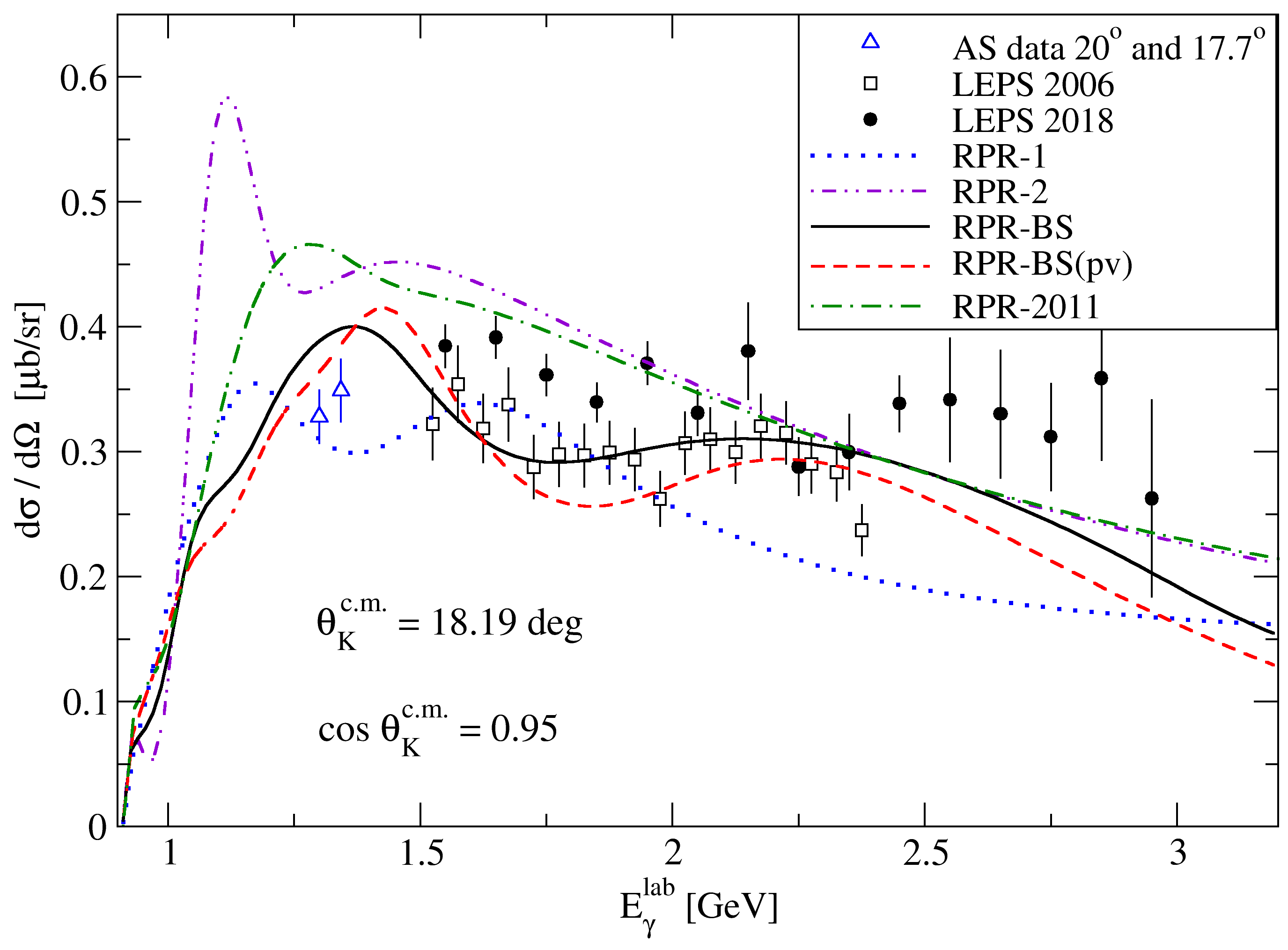}
    \caption{Differential cross section of $p(\gamma,K^+)\Lambda$ for $\cos\theta_K^{c.m.}=0.95$ is shown. We compare the model predictions with LEPS data~\cite{LEPS06,LEPS18} and two data from Ref.~\cite{AS}. }
    \label{fig:crs-leps}
\end{figure}
%
%
\begin{figure*}[ht]
    \centering
    \includegraphics[width=0.75\textwidth]{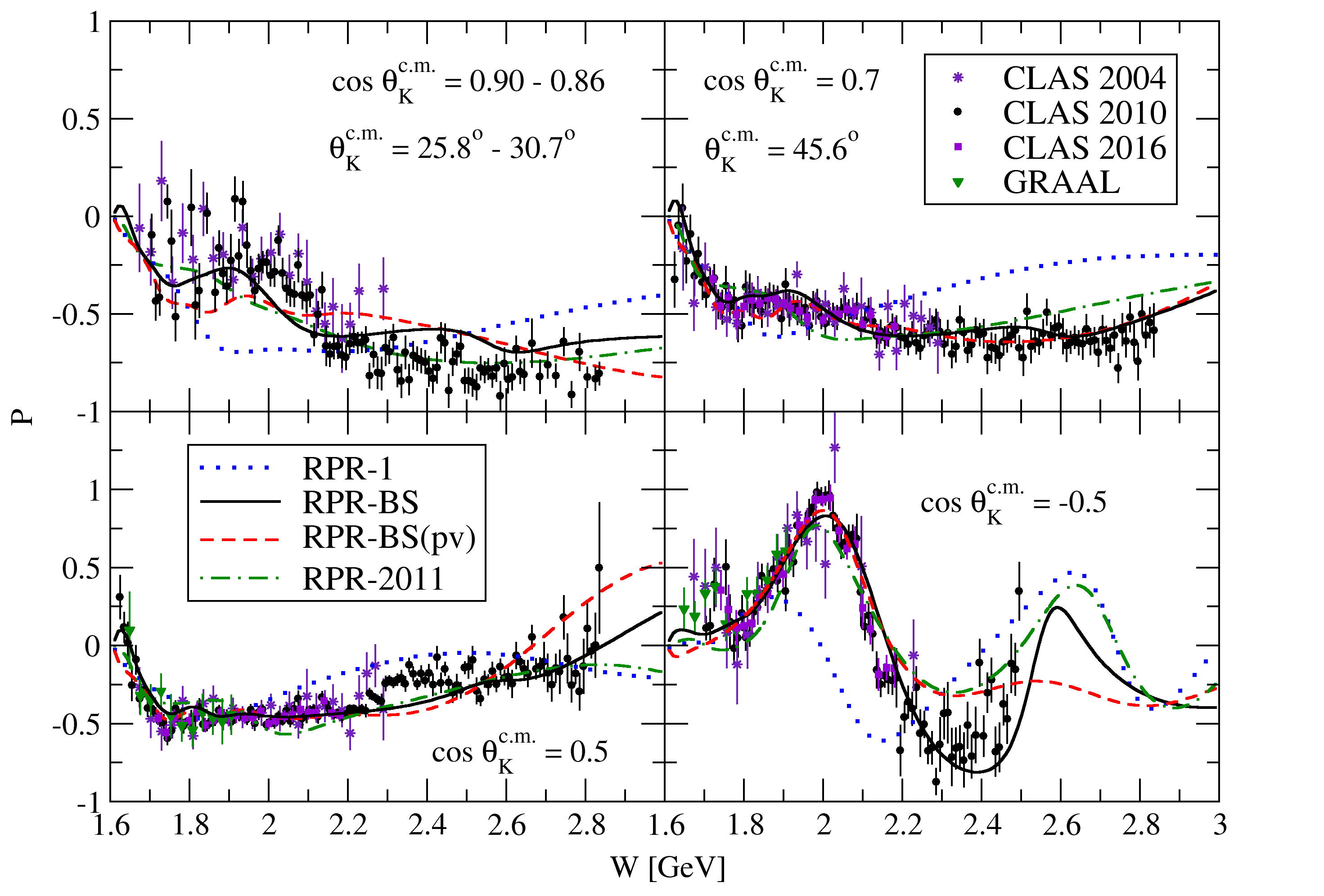}
    \caption{Results for the energetic dependence of the hyperon polarization $P$ for several kaon center-of-mass angles. Data are from the CLAS~\cite{CLAS10,CLAS04,CLAS16} and GRAAL~\cite{GRAAL07} collaborations. }
    \label{fig:pol-w}
\end{figure*}
%
%
\begin{figure*}[ht]
    \centering
    \includegraphics[width=0.75\textwidth]{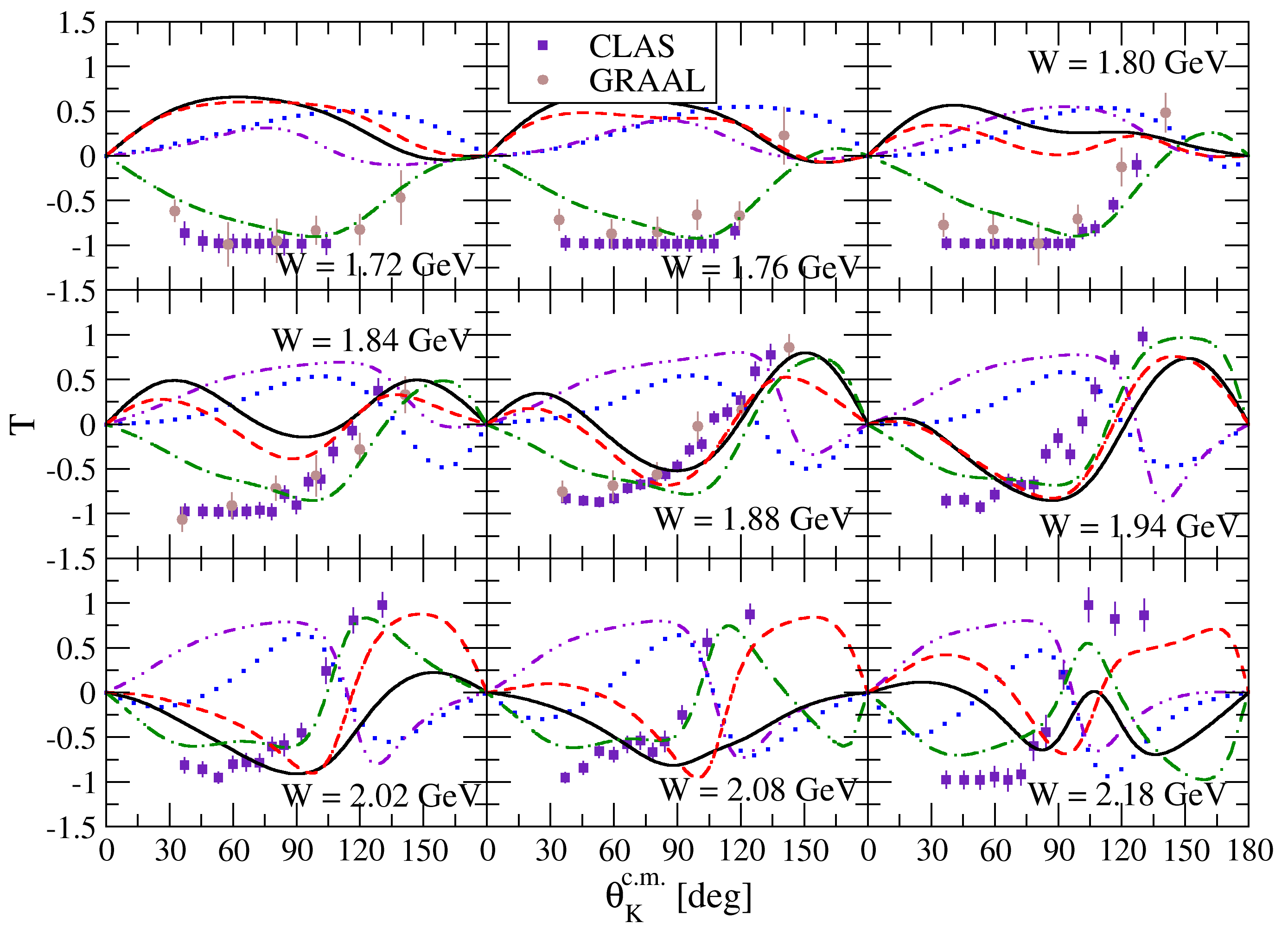}
    \caption{Predictions of the target asymmetry $T$ for several values of energy $W$. The data originate from the CLAS~\cite{CLAS16} and GRAAL~\cite{GRAAL09} experiments and notation of the curves is the same as in the Fig.~\ref{fig:crs-th0}.}
    \label{fig:t-ct0}
\end{figure*}
%
%
\begin{figure*}[ht]
    \centering
    \includegraphics[width=0.75\textwidth]{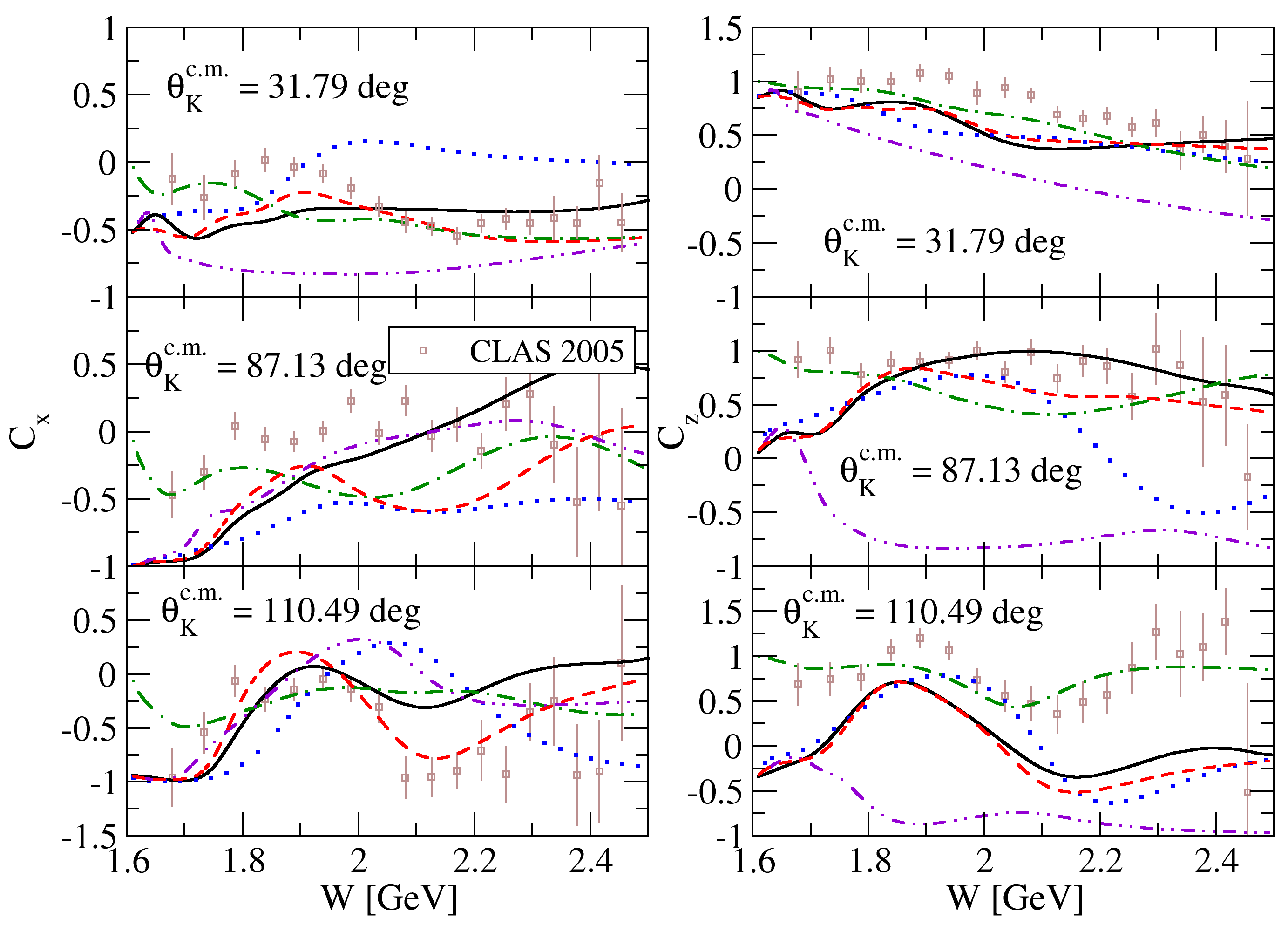}
    \caption{Predictions of the double-polarization observables $C_x$ and $C_z$ for several kaon center-of-mass angles. The data stem from the CLAS~\cite{CLAS-double} experiments and notation of the curves is the same as in the Fig.~\ref{fig:crs-th0}.}
    \label{fig:cxcz}
\end{figure*}

%
%
\begin{figure*}[ht]
    \centering
    \includegraphics[width=0.75\textwidth]{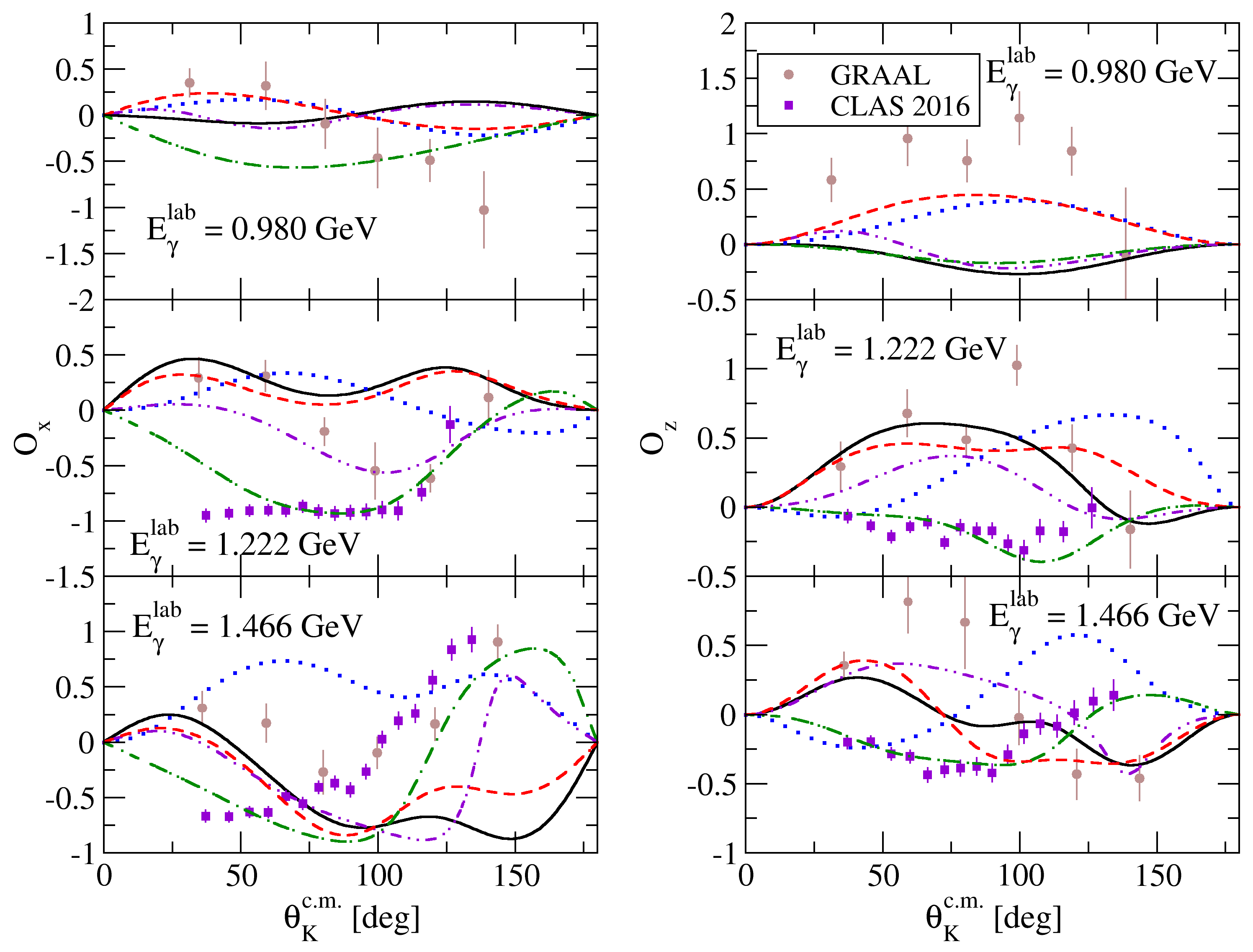}
    \caption{Predictions of the double-polarization observables $O_x$ and $O_z$ for several kaon center-of-mass angles. The data originate from the GRAAL~\cite{GRAAL09} and CLAS~\cite{CLAS16} experiments and notation of the curves is the same as in the Fig.~\ref{fig:crs-th0}.}
    \label{fig:oxoz}
\end{figure*}

We do not include an anomalous magnetic coupling in the proton exchange proportional to $\sigma^{\mu\nu}$ because of the duality hypothesis according to which only all $s$-channel or all $t$-channel poles can be included~\cite{RPR}. A combination of both $s$- and $t$-channel contributions may lead to double counting of poles. Since in the Regge model we take into account all $t$-channel poles, the amount of additional poles in the $s$ channel should be reduced to minimum. When we, however, introduce this anomalous term into the RPR-BS model we observe a suppression of cross-section predictions in the hemisphere of forward angles and an increase at backward angles. 

One of the troubling ambiguities when describing the $K^+\Lambda$ photoproduction with help of effective models is an accurate selection of the hadron form factor accounting for the extended structure of hadrons. In the robust analysis of Ghent group, they selected the multi-dipole-Gaussian shape as the most suitable one. However, we reveal in our analysis that the inclusion of this kind of hadron form factor leads to a higher $\chi^2$ value. On the other hand, opting for hadron form factors whose functional dependence on the cut-off value is much \textcolor{blue}{weaker}, \emph{i.e.} dipole or Gaussian ones, leads to unacceptable behaviour beyond the resonance region, where these form factors are not able to tame the high-spin $N^*$'s sufficiently, which leads to a rapidly soaring cross-section prediction. No matter what the cutoff parameter is, dipole and Gaussian form factors work well only in the resonance region. Therefore, we turned to a multi-dipole shape, Eq.~(\ref{eq:Fmd}), of the form factor \textcolor{blue}{for both the proton ($f_s$ with $\Lambda_{bgr}$) and $N^*$ exchanges (with $\Lambda_N$) in the $s$ channel}
which, with a reasonably small cutoff value, works well in both worlds.



In order to illustrate the roles played by particular nucleon resonances, we include a prediction of the total cross section by the RPR-BS model in Fig.~\ref{fig:totcrs}. In the upper part of this figure, where contributions of specific parts of the amplitude are shown, a noticeable feature is that the model is dominated by background (dashed line), where the \textcolor{blue}{proton exchange plays a predominant role with a tangible contribution of the contact term for $E_\gamma^{lab}<2\,\text{GeV}$} (difference between the dashed and dotted lines), and the $N^*$'s presence is tangible only right above the threshold and around the peak at $E_\gamma^{lab}\approx 1.4\,\text{GeV}$. As for the $N^*$'s contributions, the most notable ones are the destructive interferences of $S_{11}(1535)$ (above the threshold) and $F_{15}(1680)$ and $F_{15}(1860)$ (in the higher-energy domain). A most constructive interference, on the other hand, comes from the $S_{11}(1650)$ contribution above the threshold.

As we see that the contact current plays a somewhat important role in the background contribution to the total cross section, we should comment on how to interpret the role the contact term plays. The contact term mimics the contributions beyond the tree level since it can be understood as a contribution from the final-state interaction which is necessary to preserve gauge invariance~\cite{Wang}. A large contribution of the contact term, therefore, might indicate that the final-state interactions play a role, which is what we see here, but it is hard to say how big the role precisely is. What is more, the introduction of contact term can also provide us with an added flexibility in the fits to experimental data~\cite{Ronchen}.

In Figs.~\ref{fig:crs-th0}--\ref{fig:oxoz} the results and predictions of our new models, RPR-BS and RPR-BS(pv), are compared with RPR-1 and RPR-2 models which are our older fits to data~\cite{RPR-12}, where a GLV method~\cite{Guidal} for gauge-invariance restoration is used and which are motivated by the Ghent RPR-2011B model~\cite{RPR11}, and with the Ghent RPR-2011 model~\cite{DeCruz}. For obtaining results of the RPR-2011 model we made use of the online interface in Ref.~\cite{StrangeCalc}. The resulting angular dependence of the cross section is compared with experimental data in Fig.~\ref{fig:crs-th0} for six various energies in and above the resonance region. \textcolor{blue}{Whereas the models reproduce the experimental data and are also in mutual agreement in the $30^\circ < \theta_K^{c.m.} < 120^\circ$ angle region, as was found also, \emph{e.g.}, in Ref.~\cite{Kamano},} they differ particularly in the region of kaon angles $\theta_K^{c.m.}$ smaller than $30^\circ$ where there are no data currently available. The Ghent RPR-2011 model together with the RPR-2 model give the largest cross section values at zero kaon angle and then steeply decrease. Note also that the RPR-2 model was adjusted to experimental data only for $\theta_K^{c.m.}<90^\circ$, which allows for its remarkable behaviour at backward angles (most striking for $W<2\,\text{GeV}$). Our new models, on the other hand, are a bit more moderate in their forward-angle predictions which are approximately $1\,\mu\text{b/sr}$ below the RPR-2011 model at zero kaon angle and subsequently decrease more gradually. The RPR-BS model even produces a plateau-like behaviour above the resonance region ($W>2.7\,\text{GeV}$). In this model, the background terms, with a significant support from the contact current, contribute most significantly at kaon angles approximately from $30^\circ$ to $90^\circ$ and thus create the structure at around $\theta_K^{c.m.}=30^\circ$. However, their contribution for $\theta_K^{c.m.}<30^\circ$ is negligible and therefore the strength in this region comes from $N^*$'s contributions; generally, the higher the spin of the resonance the higher its contribution in forward regions but $N^*(5/2)$ are also active at backward angles (see \emph{e.g.} the peak around $\theta_K^{c.m.}=140^\circ$ at $W=2.105\,\text{GeV}$). Interestingly, when we assume a pseudovector coupling in the $K\Lambda N$ vertex the behaviour of \textcolor{blue}{the model changes slightly as it produces a plateau in the forward kaon angles even at energies $W>2.5\,\text{GeV}$. However, the dynamics of the model now differs: In comparison with the RPR-BS model the background contribution in the RPR-BS(pv) model is much smaller, even though it is strong enough to produce a structure around $\theta_K^{c.m.}\approx 30^\circ$ visible at higher energies, and, thus, the contribution of $N^*$'s is more pronounced, particularly at forward angles.}

The model behaviour at forward angles seems to be strongly influenced by the choice of gauge-invariance restoration method. In the GLV procedure for gauge-invariance restoration, which is implemented in our older RPR-1 and RPR-2 models, the crucial contribution at forward angles stems from the proton exchange, which is governed by $g_{K\Lambda N}$ coupling constant. Generally, the smaller is the coupling constant the higher is the cross section at forward angles. In the RPR-1 model, this coupling parameter acquires a value of $-1.45$, whereas in the RPR-2 model the $g_{K\Lambda N}$ is much lower, specifically $-3.00$. The proton exchange contribution in the RPR-2 model is thus much stronger at very small kaon angles. An interested reader may find a much more thorough analysis of this topic in Ref.~\cite{DSphd}.

The energetic dependence of the cross section is shown in the Fig.~\ref{fig:crs-w} for four angles in the forward hemisphere. The data reveal a two-peak structure and so do the models as all of them are in accordance with experimental data except for the sharp peak right above the threshold at $\cos\theta_K^{c.m.}=0.8$ in the CLAS 2010 data set which is apparently excluded also by the older CLAS 2005 measurement. Most probably, the $S_{11}(1650)$ state, together with its constructive interference with other terms, plays the decisive role in creating or not creating that sharp a peak in model predictions. In the RPR-2 model, coupling parameters of this state are \textcolor{blue}{more than three times} as large as in other RPR models of ours and an order of magnitude higher in comparison with our isobar models. No wonder, then, that this peak is formed in the RPR-2 model and not in the other models. The description of the cross section by the RPR-BS model above the threshold is shaped predominantly by an interplay of $S_{11}(1535)$ and $S_{11}(1650)$ states while the dip around $W\approx 1.9\,\text{GeV}$ is modelled by $D_{13}(1875)$ which interferes destructively with other terms. Above 2~GeV, the main (destructive) contributions come from $F_{15}(1680)$ and $F_{15}(1860)$ and they decrease with kaon angle. Even though these spin-5/2 resonances resonate higher than where their masses are, they eventually get suppressed so that the high-energy region is completely modelled by the Regge-like background. This aptly illustrates the sufficiency of the multi-dipole hadron form factor for taming even the high-spin $N^*$'s contributions. \textcolor{blue}{The most notable change in the behaviour of $N^*$ states in the RPR-BS and RPR-BS(pv) models gives the $S_{11}(1535)$ state which has a different sign in the RPR-BS(pv) model than in the RPR-BS model. Its contribution in the RPR-BS model shows a destructive interference with other terms while in the RPR-BS(pv) model there is a constructive interference.}

As mentioned in the Introduction, a realistic behaviour of the $\Lambda$-production amplitude is vital for obtaining reliable  predictions for the hypernucleus-production cross section~\cite{hypernucleus},  which is substantial only in the region of very small kaon angles. Therefore, we also examined model predictions in this kinematic region, \textcolor{blue}{see Figs.~\ref{fig:crs-6deg} and~\ref{fig:crs-leps}}.
In the very forward-angle region with $\theta_K^{c.m.}<20^\circ$, the experimental data are highly insufficient, which leaves us with predictions of our models, see Fig.~\ref{fig:crs-6deg}. We can, however, get a helpful hint on how the model prediction should behave from one photoproduction datum by Bleckmann {\it et al.}~\cite{Bleck}, two electroproduction data points~\cite{Brown,E94-107} with a very small virtual-photon mass, and from our knowledge from hypernuclei calculations that the center-of-mass cross section at $\theta_K^{c.m.}=2^\circ$ and around $W=2.2\,\text{GeV}$ should be at least $0.4\,\mu\text{b/sr}$ \cite{hypernucleus}. See also the discussion on the elementary reaction in Garibaldi {\it et al.}~\cite{hypernucleus}. Complying with the latter condition in particular helped us in selecting the final models from the plenty of fits. While the RPR-2011 model predicts a predominantly structureless cross section, our models reveal some resonant features, see Fig.~\ref{fig:crs-6deg}. The first peak in our new models is most probably the result of an interplay of many nucleon states, where the $S_{11}(1535)$ and $S_{11}(1650)$ are the most prominent ones. The second peak is then created by spin-5/2 $F_{15}(1680)$ and $F_{15}(1860)$ states, which is in accord with what we stated above about the magnitude of predictions of high spin resonances in the forward-angle kinematic region. \textcolor{blue}{The RPR-BS and RPR-BS(pv) models give almost the same shape of the cross section and their predictions differ only in magnitude. One can find an interesting hint of a peak right at the threshold, which is created by a strong contribution of $S_{11}(1650)$ resonance. This is quite a remarkable feature, as other models either reveal a similar peak at slightly higher energies (RPR-2) or they do not present it at all (it probably gets smoothed out by interfering with other terms).} The older RPR models of ours give predictions that do not differ much in shape but they differ in magnitude. We have already discussed the role the proton exchange plays at very forward angles but let us point out that another difference between RPR-1 and RPR-2 models is a cut-off parameter value for the hadron form factor of nucleon resonances (please note that there is no hadron form factor introduced for background terms in the GLV gauge-invariance restoration scheme). In the RPR-1 model the hadron form factor with a cutoff value 2~GeV suppresses nucleon resonances efficiently, which results in recession above $E_\gamma^{lab}=2\,\text{GeV}$. On the other hand, the RPR-2 model prediction diminishes slowly as the cutoff value for its hadron form factor is 3~GeV. Both RPR-1 and RPR-2 models exploit the multidipole-Gaussian shape of hadron form factor.
From Fig.~\ref{fig:crs-6deg} and its discussion one can see that in the very forward angle region the models are still unconstrained by the data. Figure~\ref{fig:crs-6deg} hence collects mere predictions of various models and as such reveals diverse forms of dynamics which strongly affect the hypernucleus production results. 

\textcolor{blue}{Another figure which illustrates the varying predictions of various models is the Fig.~\ref{fig:crs-leps}. The shapes of model predictions are analogous to the ones in the Fig.~\ref{fig:crs-6deg} but this figure is replenished by a number of experimental data, namely the LEPS 2006~\cite{LEPS06} and LEPS 2018~\cite{LEPS18} data and the data collected in Ref.~\cite{AS}. One can readily see that our new models are in concert with the LEPS 2006 data, whereas they underpredict the LEPS 2018 data which are also well above the older LEPS data. Since we include both LEPS data sets in the fits, this discrepancy may lead to an increased $\chi^2$ value of our models. Both peaks in RPR-BS model predictions are created by nucleon resonances; the first one mainly by $D_{13}(1700)$ and the other one is a result of an interplay of $F_{15}(1860)$ and $F_{15}(2000)$. The most notable contribution to the first peak in the RPR-BS(pv) model comes from the $D_{13}(1700)$ and $P_{13}(1720)$ states and the broad peak around $E_\gamma^{lab}\approx 2.2\,\text{GeV}$ is created by $F_{15}(2000)$ and $D_{15}(2060)$. In the kinematic region shown, the background terms of RPR-BS model create a very broad peak at around $E_\gamma^{lab}\approx 1.2\,\text{GeV}$, which steadily decreases at higher energies, and contributions of both Regge trajectories increase smoothly with energy but are not larger than $0.05\,\mu\text{b/sr}$, similarly to the whole background in the RPR-BS(pv) model (\emph{i.e.} Regge trajectories dominate the background in the RPR-BS(pv) model at forward kaon angles).}

In Fig.~\ref{fig:pol-w}, there are results for hyperon polarization $P$ for several kaon angles. \textcolor{blue}{Our new  models are in a good agreement with data in all kinematic regions shown. The only exception may be the upper left part of Fig.~\ref{fig:pol-w} with $\cos\theta_K^{c.m.}=0.9-0.86$ where the actual shape of hyperon polarization near the threshold is hard to guess, thanks to considerable inconsistencies in data.} The dominant contributions to $P$ in the forward-angle hemisphere come from background and higher-spin nucleon resonances whereas the contact term contributes at central angles mainly and the presence of spin-1/2 nucleon resonances is noticeable only at the threshold area. The role of nucleon states then lies especially in interfering among themselves and other terms and thus creating the subtle shapes as both spin-3/2 and spin-5/2 nucleon resonances on their own produce shapes which are far from what can be seen in the result of the complete model. The RPR-2011 model, \textcolor{blue}{as well as our fits}, is able to capture the hyperon-polarization data also at very forward angles and at $\cos\theta_K^{c.m.}=-0.5$ in the transition between resonant and high-energy regions produces a structure according to experimental data. \textcolor{blue}{In case of the RPR-BS model, this structure is created predominantly by the $D_{15}(2570)$ state and without this resonance we get a plateau at a value of $-0.7$. Similarly, in the RPR-BS(pv) model we observe a considerable contribution at $W>2.2\,\text{GeV}$ from the $D_{15}(2060)$, which is unfortunately not as strong as the $D_{15}(2570)$ in the RPR-BS model and thus the RPR-BS(pv) model fails to reproduce the shape of data in this region.} As we pointed out in the preceding section, the CLAS 2016~\cite{CLAS16} data are in concert with the older CLAS data and also with the data from the GRAAL facility in Grenoble~\cite{GRAAL07}.

In Fig.~\ref{fig:t-ct0}, we show predictions of our models for the target asymmetry $T$ and compare them with CLAS~\cite{CLAS16} and GRAAL~\cite{GRAAL09} data which are mutually well consistent. We can very roughly say that the higher the energy \textcolor{blue}{and the larger the kaon angle}, the closer our models, particularly the one with the pseudovector coupling in the $K\Lambda N$ vertex, come to the data, even though their parameters were not adjusted to the target asymmetry. The only model which can capture the shape of data at all energies shown is the Ghent RPR-2011 model.

During the fitting procedure, we did not fit to experimental data on double-polarization observables $C_x$, $C_z$, $O_x$, and $O_z$ and hence Figs.~\ref{fig:cxcz} and \ref{fig:oxoz} show mere predictions of the models. For $C_x$ we have a reliable prediction by the RPR-BS model for all angles shown except for the threshold region at $\theta_K^{c.m.}=87.13^\circ$, where the model underpredicts the data, and at $\theta_K^{c.m.}=110.49^\circ$ for $W>2\,\text{GeV}$, where its predictions do not lie within the data error bars. The RPR-BS(pv) model with pseudovector coupling in the $K\Lambda N$ vertex works similarly \textcolor{blue}{but gives better predictions at $\theta_K^{c.m.}=110.49^\circ$} and the Ghent RPR-2011 model shows slightly less structures than what we see in the data. It is the RPR-2011 model which captures the $C_z$ data at high angles and high energies best while our models fail to reproduce even the shape of data. However, they give good predictions of $C_z$ in the forward hemisphere of kaon angles. The agreement with $O_x$ and $O_z$ data is much worse since in many cases the models predict structures with opposite sign in comparison with the \textcolor{blue}{CLAS 2016~\cite{CLAS16}} data or their predictions lack any structure \textcolor{blue}{shown in this data set}. Generally, we observe more structures in the model predictions at higher energies and for $E_\gamma^{lab}=1.222$~GeV our RPR models can at least capture the shape of the GRAAL~\cite{GRAAL07} data. Unfortunately, we also see that the recent CLAS~\cite{CLAS16} data strictly oppose the older GRAAL data in some kinematic regions, which makes the analysis much more precarious.

\begin{figure}[h]
    \centering
    \includegraphics[width=0.50\textwidth]{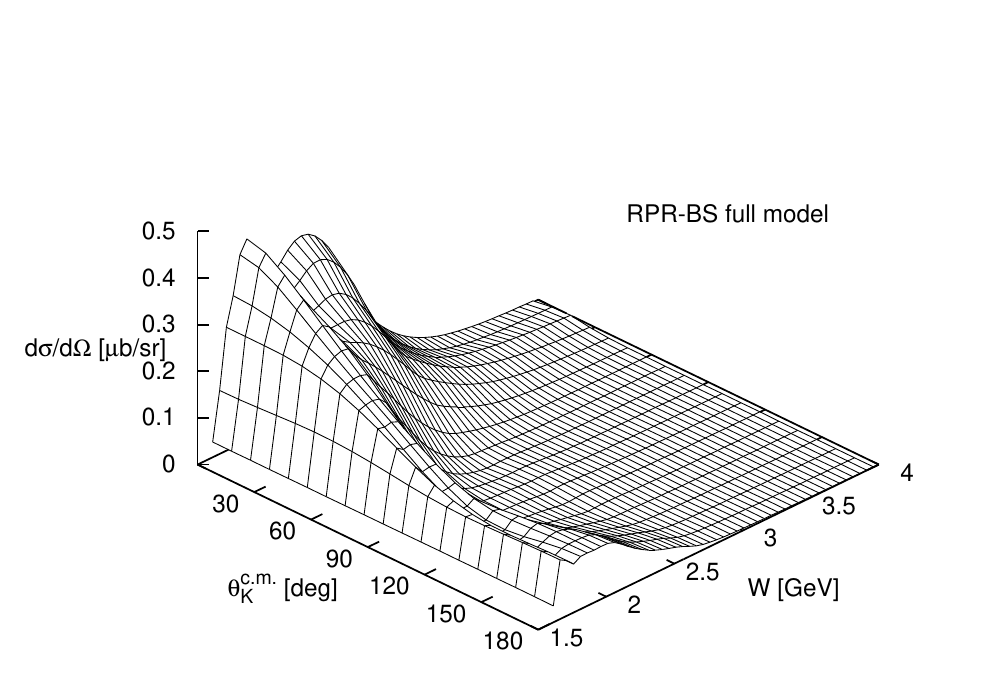}
    \caption{Overall description of the cross section for the $p(\gamma,K^+)\Lambda$ reaction by the RPR-BS model from the threshold up to $4\,\text{GeV}$ and for all kaon angles.}
    \label{fig:RPR-BS-3D}
\end{figure}

\begin{figure}[h]
    \centering
    \includegraphics[width=0.50\textwidth]{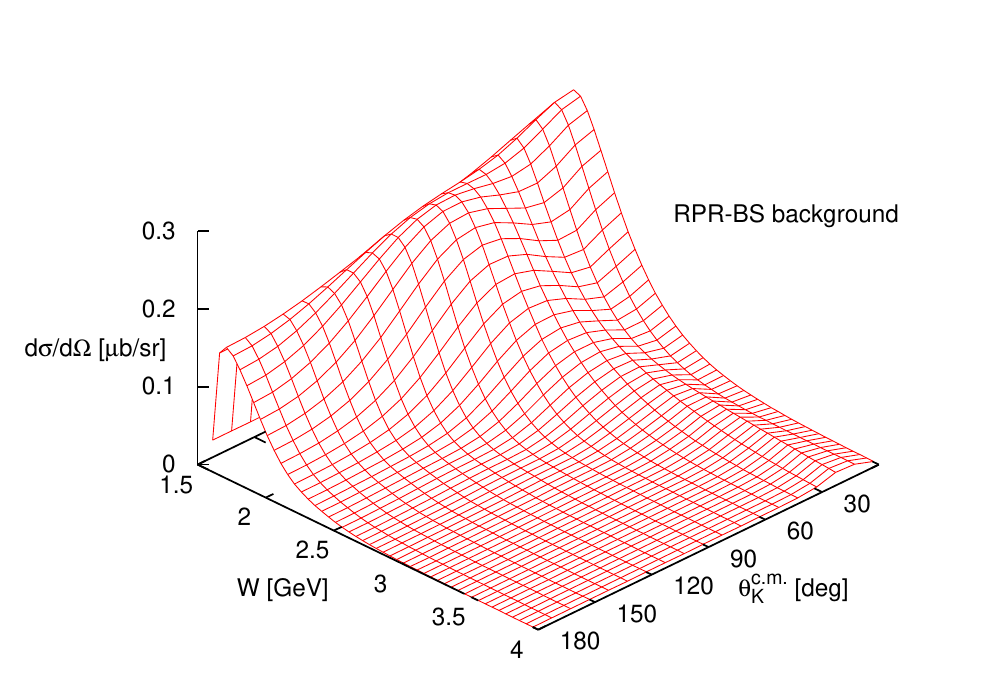}
    \caption{The same as Fig.~\ref{fig:RPR-BS-3D} but this time with background terms only.}
    \label{fig:RPR-BS-3D-bgr}
\end{figure}

\begin{figure}[h]
    \centering
    \includegraphics[width=0.50\textwidth]{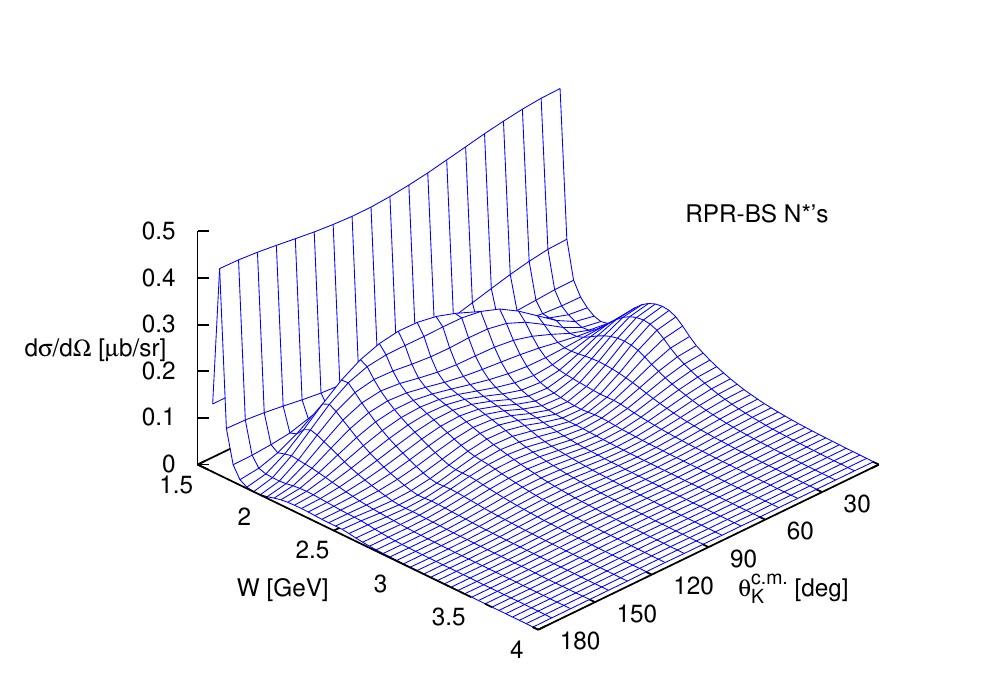}
    \caption{The same as Fig.~\ref{fig:RPR-BS-3D} but with nucleon resonances solely.}
    \label{fig:RPR-BS-3D-N}
\end{figure}

In Figs.~\ref{fig:RPR-BS-3D}, \ref{fig:RPR-BS-3D-bgr}, and \ref{fig:RPR-BS-3D-N}, we show an overall description of the cross section for the  $p(\gamma,K^+)\Lambda$ production process by the RPR-BS model, by the mere background terms, and by its set of nucleon resonances, respectively, for all angles and for energies from the threshold to 4~GeV, \emph{i.e.} well beyond the resonance region. We can see that at higher energies, there is some strength only at very forward angles, $\theta_K^{c.m.}<30^\circ$, which is apparently caused by the background terms. The nucleon states do not contribute anywhere \textcolor{blue}{well above the resonance region, \emph{i.e.} for $W>3\,\text{GeV}$,} as requested, and the high-energy region is therefore described by the Regge background only. This shows beyond any doubt that even the multi-dipole hadron form factor can suppress contributions of nucleon resonances sufficiently so that they vanish at the edge of the resonance region. \textcolor{blue}{In Fig.~\ref{fig:RPR-BS-3D-N}, which collects the contributions of the $N^*$'s, the sharp peak at the threshold which almost does not depend on kaon angle is produced mainly by the $N^*(1/2)$ states, the peak around $W=2\,\text{GeV}$ is shaped above all by the $N^*(3/2)$ states and the other structures are the result of interference among $N^*(3/2)$ and $N^*(5/2)$ states.} The complicated shape produced in the resonance region therefore seems to be a result of a rather intricate interference among many $N^*$ and background terms. In the forward angles, there appears to be constructive interference producing the second peak around approximately 2~GeV, whereas we surely observe a destructive interference at backward angles leading to a suppressed cross section.

%
%

%
%

%
%

\section{Conclusion}
\label{sec:con}
We have constructed a new version of the Regge-plus-resonance model for $p(\gamma,K^+)\Lambda$ utilizing a new method to maintain  gauge invariance that is based on generalized Ward-Takahashi indentities. In this method a Reggeized contact term is included together with the proton exchange in the $s$ channel with the standard Feynman propagator. Another novel feature of the model is the presence of the hadron form factor in the proton exchange that constitutes an important contribution to the non resonant part of the photoproduction amplitude. The nucleon resonances with higher spins are treated in the frame of consistent formalism that we used also in  our recent isobar models. In the analysis we considered both pseudoscalar and pseudovector forms of couplings in the strong $K\Lambda N$ vertex.
Parameters of the model were adjusted to ample data in the resonance region and to available CLAS and LEPS data above this region with \textcolor{blue}{$\chi^2= 1.69$ and 1.74 for the model with the pseudoscalar and pseudovector coupling in the $K\Lambda N$ vertex, respectively.}  A set of nucleon resonances contributing significantly to the process was carefully selected and some resonance parameters (mass and width) were gently modified. The chosen  set of $N^*$'s agrees quite well with that selected in the Ghent analysis. Concerning the fitting of parameters, let us note that including all spin observables in the data set and using a more sophisticated method (such as the one recently used in Ref.~\cite{Landay}) would probably lead to an even more satisfactory result. Performing such a robust analysis,  however, was beyond the scope of this work but it might be one of the objectives of our future research.

Satisfactory description of the cross sections and polarizations was achieved. However, the model predictions still diverge for very small kaon angles. In the new RPR model the background part is dominated by the contact term, which mimics the higher-order effects, pointing out to importance of the final-state interactions.

\section*{Acknowledgements}
The authors thank Helmut Haberzettl for useful discussions and careful reading of the manuscript. This work was supported by the Czech Science Foundation GACR Grant No. 19-19640S and by the JSPS Grant No.~16H03995.

\appendix

\begin{widetext}

\section{Contact current contributions}
In the case of pseudoscalar coupling in the strong vertex, the contact term contribution to scalar amplitudes defined in Ref.~\cite{SB16} reads
\begin{subequations}
\begin{align}
\mathcal{A}_2 = {}& -2 eg_{K\Lambda p} \Bigg\{  \frac{\mathcal{F}_t-1}{t-m_K^2}f_s + \frac{f_s -1}{s-m_p^2} \mathcal{F}_t + \hat{A}(s,t,u)(1-f_s)(1-\mathcal{F}_t)\left[ \frac{1}{t-m_K^2} + \frac{1}{s-m_p^2} \right] \Bigg\},
\\
\mathcal{A}_3 = {}& 2eg_{K\Lambda p} \Bigg\{ \frac{\mathcal{F}_t -1}{t-m_K^2}f_s - \hat{A}(s,t,u)(1-f_s)(1-\mathcal{F}_t)\frac{1}{t-m_K^2} \Bigg\}.
\end{align}
\label{eq:cc-ps-amplitudes}
\end{subequations}

When we assume a pseudovector coupling in the strong vertex, the contact term contribution, beyond the contributions of Eqs. (\ref{eq:cc-ps-amplitudes}), reads

\begin{subequations}
\begin{align}
\mathcal{A}_4 = {}& 2eg^\prime_{K\Lambda p} \left\{  \frac{\mathcal{F}_t-1}{t-m_K^2}f_s + \frac{f_s-1}{s-m_p^2}\mathcal{F}_t - \hat{A}(s,t,u)(1-f_s)(1-\mathcal{F}_t)\left[ \frac{1}{t-m_K^2} + \frac{1}{s-m_p^2} \right] \right\},\\
\mathcal{A}_5 = {}& -2eg^\prime_{K\Lambda p} \left\{ \frac{\mathcal{F}_t-1}{t-m_K^2}f_s + \hat{A}(s,t,u)(1-f_s)(1-\mathcal{F}_t)\frac{1}{t-m_K^2} \right\},\\
\begin{split}
\mathcal{A}_6 = {}& 2eg^\prime_{K\Lambda p} \bigg\{ \frac{f_s -1}{s-m_p^2}\mathcal{F}_t \frac{k\cdot p}{k^2} - \frac{\mathcal{F}_t-1}{t-m_K^2}f_s \frac{1}{k^2}(k\cdot p_\Lambda - k\cdot p) \\ 
& + \hat{A}(s,t,u)(1-f_s)(1-\mathcal{F}_t)\left[ \frac{1}{t-m_K^2}\frac{1}{k^2}(k\cdot p_\Lambda - k\cdot p) - \frac{1}{s-m_p^2}\frac{k\cdot p}{k^2} \right] \bigg\}.
\end{split}
\end{align}
\end{subequations}

Electric part of the Born $s$-channel contribution with the pseudovector coupling can be recast to the compact form
\begin{equation}
\begin{split}
\mathbb{M}_{Bs-el}^{(PV)} & = \bar{u}(p_\Lambda)(-eg_{K\Lambda p}^\prime) f_s \not\! p_K\gamma_5 \frac{\not\!p +\not\! k + m_p}{s-m_p^2}\gamma_\mu\varepsilon^\mu u(p) \\
& = \bar{u}(p_\Lambda)\gamma_5 \frac{eg^\prime_{K\Lambda p}}{s-m_p^2} f_s \bigg[ (m_\Lambda+m_p)(\mathcal{M}_1+2\mathcal{M}_2) - (1+2k\cdot p/k^2)\mathcal{M}_6 + (s-m_p^2) (m_\Lambda+m_p+\not\! k)\frac{k\cdot \varepsilon}{k^2} \bigg]u(p).
\end{split}
\label{eq:s-pv}
\end{equation}
Scalar amplitudes resulting from the electric part of the Born $s$ channel then are
\begin{equation}
\mathcal{A}_1 = f_s \frac{eg_{K\Lambda p}}{s-m_p^2} = \frac{1}{2}\mathcal{A}_2, \,\,\, \mathcal{A}_6 = -f_s \frac{eg^\prime_{K\Lambda p}}{s-m_p^2}\left(1+2\frac{k\cdot p}{k^2}\right).
\end{equation}

Magnetic part of the Born $s$-channel contribution in the pseudovector coupling
\begin{equation}
\mathbb{M}_{Bs-mg}^{(PV)} = \bar{u}(p_\Lambda) (-eg^\prime) f_s \not\!p_K \gamma_5 \frac{\not\!p +\not\! k + m_p}{s-m_p^2} i\frac{\kappa_p}{2m_p} \sigma_{\mu\nu} k^\nu \varepsilon^\mu u(p)
\end{equation}
can be recast into the compact form
\begin{equation}
\mathbb{M}_{Bs-mg}^{(PV)} = \bar{u}(p_\Lambda) \gamma_5 \frac{eg^\prime}{s-m_p^2}f_s \frac{\kappa_p}{2m_p}[ (2 k\cdot p+k^2)\mathcal{M}_1 + 2(m_\Lambda+m_p)\mathcal{M}_4 - (m_\Lambda+m_p)\mathcal{M}_6 ] u(p),
\end{equation}
from which one can extract the scalar amplitudes
\begin{equation}
\mathcal{A}_1 = \frac{eg^\prime}{s-m_p^2}f_s \frac{\kappa_p}{2m_p}(2 k\cdot p + k^2),\,\,\,
\mathcal{A}_4 = \frac{eg}{s-m_p^2}f_s \frac{\kappa_p}{m_p} = -2\mathcal{A}_6.
\end{equation}

Born $t$-channel contribution with the pseudovector coupling in the strong vertex can be recast into the compact form
\begin{equation}
\begin{split}
\mathbb{M}_{Bt}^{(PV)} & = \bar{u}(p_\Lambda)(-eg^\prime_{K\Lambda p})\mathcal{F}_t (\not\! p \,\,- \not\! p_\Lambda) \gamma_5 \frac{(2p_K-k)_\mu}{t-m_K^2}\varepsilon^\mu u(p)\\
& = \bar{u}(p_\Lambda)\gamma_5 eg_{K\Lambda p} \frac{\mathcal{F}_t}{t-m_K^2}\bigg[ 2(\mathcal{M}_2-\mathcal{M}_3) - (t-m_K^2) \frac{k\cdot \varepsilon}{k^2}  \bigg]u(p).
\label{eq:t-pv}
\end{split}
\end{equation}
Scalar amplitudes of this contribution then read
\begin{equation}
\mathcal{A}_2 = 2 \mathcal{F}_t \frac{eg_{K\Lambda p}}{t-m_K^2} = - \mathcal{A}_3.
\end{equation}

\end{widetext}

\section{Regge Trajectories and Propagators}
\label{App:Regge}

At the energies of a few GeV and higher, where no individual resonances can be distinguished, the dynamics of the process is governed by the exchange of $t$-channel Regge trajectories. This choice is motivated by the shape of the $K^+\Lambda$ photoproduction cross section which is peaked on small $|t|$, \emph{i.e.} on small kaon angles $\theta_{K}^{c.m.}$. This behaviour indicates a dominant role played by $t$-channel kaon exchanges.

\begin{figure}[t]
\centering
\includegraphics[width=0.51\textwidth]{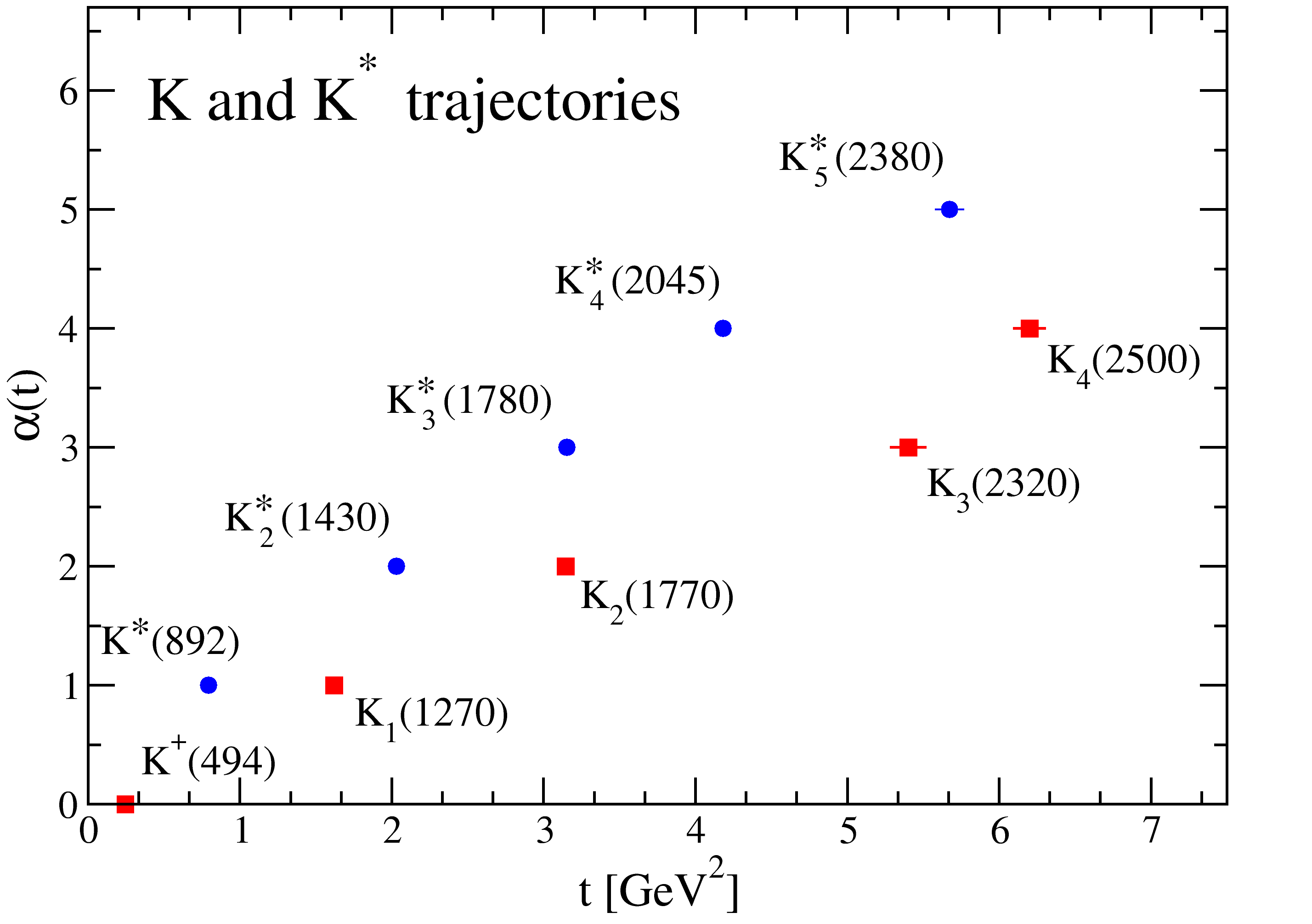}
\caption{Chew-Frautschi plot for the two lightest kaon trajectories assumed in our analysis. The squares and dots represent trajectories with parity $+1$ and $-1$, respectively. Both trajectories are linear to a very good approximation.}
\label{fig:regge-traj}
\end{figure}

The Regge trajectories, which are often called after a lightest member (so-called first materialization) of the particular trajectory, connect spin and mass squared of the exchanged particle. When the spins of a set of resonant states are plotted against their mass squared in a Chew-Frautschi plot, see Fig.~\ref{fig:regge-traj}, it is observed that all Regge trajectories can be reasonably well parameterized by means of a linear function
\begin{equation}
\alpha_X(t)=\alpha_{X,0}+\alpha'_X(t-m_X^2),
\end{equation}
with $m_X$ and $\alpha_{X,0}$ the mass and spin of the trajectory lightest member $X$, respectively. What is more, $\alpha'_X$, which is the slope of the trajectory, happens to be close to an universal constant for all trajectories and acquires the value of $0.8\,\mbox{GeV}^2$. Trajectory equations for $K^+(494)$ and $K^{*}(892)$ read
\begin{subequations}
\begin{align}
\alpha_{K(494)}(t)&=0.70\,(t-m_{K}^2),\\
\alpha_{K^{*}(892)}(t)&=1+0.85\,(t-m_{K^{*}}^2),
\end{align}
\label{eq:Regge-traj}
\end{subequations}
respectively. Note that $t=m_X^2$ can never be reached in the physical region of the process as $t$ is negative in this region.

%
%

An efficient way to model trajectory exchanges involves embedding the Regge formalism into a tree-level effective-field model. The amplitude for the $t$-channel exchange of a linear kaon trajectory $\alpha(t)$ can be obtained from the standard Feynman amplitude by replacing the usual pole-like Feynman propagator of a single particle with a Regge one of the form 
\begin{equation}
\begin{split}
\mathcal{P}_{Regge}^{\zeta=\pm 1}(s,t)={}& \left(\frac{s}{s_{0}}\right)^{\alpha(t)}\frac{\pi\alpha'}{\sin[\pi\alpha(t)]}\frac{1+\zeta \texttt{e}^{-\texttt{i}\pi\alpha(t)}}{2}\\
& \times\frac{1}{\Gamma[\alpha(t)+1]},
\end{split}
\label{eq:ReggeP}
\end{equation}
while keeping the vertex structure given by the Feynman diagrams which correspond to the first materialization of the trajectory.

While deriving the Regge propagator, one has to differentiate between two signature parts of the trajectories, $\zeta=\pm 1$, in order to obey the convergence criteria: $\zeta =+1$ corresponds with the even and $\zeta=-1$ with the odd partial waves. Thus, a summation over this factor is to be done in the propagator. Unfortunately, the theory does not allow us to determine the relative sign between the even and odd parts of the trajectory. We, therefore, end up either with a so-called constant phase, identical to 1, or a rotating phase which gives rise to a complex factor of $\texttt{exp}(-\texttt{i}\pi\alpha(t))$. As was revealed in Ref.~\cite{DeCruz}, both trajectories with rotating phases are clearly favoured by data.

In our treatment of $K^+\Lambda$ photoproduction, we identify the $K^+(494)$ and $K^{*}(892)$ trajectories as the dominant contributions to the high-energy amplitude. 
The corresponding propagators for the $K^+(494)$ and $K^{*}(892)$ trajectories have the following form~\cite{RPR}
\begin{subequations}
\begin{align}
\begin{split}
\mathcal{P}_{Regge}^{K(494)}(s,t)&=\frac{(s/s_{0})^{\alpha_{K}(t)}}{\sin[\pi\alpha_{K}(t)]}\frac{\pi\alpha'_{K}}{\Gamma[1+\alpha_{K}(t)]} \\
& \;\;\; \times \left\{
\begin{array}{c}
1 \\
\texttt{e}^{-i\pi\alpha_{K}(t)}
\end{array}
\right\},
\end{split}
\\
\begin{split}
\mathcal{P}_{Regge}^{K^{*}(892)}(s,t)&=\frac{(s/s_{0})^{\alpha_{K^{*}}(t)-1}}{\sin[\pi\alpha_{K^{*}}(t)]}\frac{\pi\alpha'_{K^{*}}}{\Gamma[\alpha_{K^{*}}(t)]} \\ 
& \;\;\; \times \left\{
\begin{array}{c}
1 \\
\texttt{e}^{-i\pi(\alpha_{K^{*}}(t)-1)}
\end{array}
\right\}.
\end{split}
\label{eq:ReggeP-K}
\end{align}
\end{subequations}

As can be seen from the definition of the Regge propagators, there are poles at non negative integer values of $\alpha_X(t)$, which correspond to the zeroes of the sine function which are not compensated by the poles of the $\Gamma$ function. Here comes the interpretation of the Regge propagator effectively incorporating the exchange of all members of the $\alpha_X(t)$ trajectory. In the physical region of the process under study (with $t<0$), these poles cannot be reached.

The separation of the Regge amplitude into two different signatures is a theoretical request to ensure convergence, experimentally both trajectories shown in (\ref{eq:Regge-traj}) coincide with one another. The residue for the lowest materialisation is, therefore, assumed to be used for the combined trajectory of both odd and even parity. This assumption is then called degeneracy. Whether a trajectory should be treated as degenerate or non degenerate, depends less on the trajectory equations themselves than on the process studied. It is the structure of the observed cross section that gives a hint whether the degeneracy is a valid supposition for a given channel or not. Non degenerate trajectories lead to peaks in the differential cross section while a smooth differential cross section indicates degenerate trajectories~\cite{Collins}. Since no obvious structure is present in the $p(\gamma, K^+)\Lambda$ cross-section data for $E_{\gamma}^{lab}\geq 4\,\mbox{GeV}$, both the $K^+(494)$ and $K^{*}(892)$ trajectories are supposed to be degenerate~\cite{RPR}.


\begin{thebibliography}{10}
\bibitem{Capstick} S.~Capstick and W.~Roberts,   Prog.~Part.~Nucl.~Phys. \textbf{45}, S241 (2001).

\bibitem{Loring} U.~Loring, B.~C.~Metsch, and H.~R.~Petry,      Eur.~Phys.~J. A \textbf{10}, 309 (2001); 
\textbf{10}, 395 (2001); \textbf{10}, 447 (2001).

\bibitem{hypernucleus} 
P. Bydžovský, M. Sotona, T. Motoba, K. Itonaga, K. Ogawa, and O. Hashimoto, 
   Nucl. Phys. A \textbf{881}, 199 (2012); F. Garibaldi \emph{et al.}, \textcolor{blue}{Phys. Rev. C {\bf 99}, 054309 (2019).}

\bibitem{ZpL95}Zhenping Li, Phys. Rev. C {\bf 52}, 1648 (1995); 
Zhenping Li, Hongxing Ye, Minghui Lu, 
Phys. Rev. C {\bf 56}, 1099 (1997).

\bibitem{LLP95}D.~Lu, R.~H.~Landau, and S.~C.~Phatak, 
Phys. Rev. C {\bf 52}, 1662 (1995). 

\bibitem{FHZ91}G.~R.~Farrar, K.~Huleihel, and H.~Zhang, 
Nucl. Phys. B {\bf 349}, 655 (1991).

\bibitem{Giessen} V. Shklyar, H. Lenske, and U. Mosel, 
Phys. Rev. C {\bf 72}, 015210 (2005); 
R. Shyam, O. Scholten, and H. Lenske, 
Phys. Rev. C {\bf 81}, 015204 (2010).

\bibitem{JDiaz} B. Julia-Diaz, B. Saghai, T.-S.H. Lee, and F. Tabakin, 
Phys. Rev. C {\bf 73}, 055204 (2006).

\bibitem{Anisovich} A.V. Anisovich, V. Kleber, E. Klempt, V.A. Nikonov, A.V. Sarantsev, and U. Thoma, 
Eur. Phys. J. A {\bf 34}, 243 (2007).

\bibitem{Borasoy} B. Borasoy, P.C. Bruns, U.-G. Meissner, and R. Nissler, 
Eur. Phys. J. A {\bf 34}, 161 (2007).

\bibitem{SB16}
D.~Skoupil, P.~Byd\v{z}ovsk\'{y}, Phys. Rev. C \textbf{93}, 025204 (2016).

\bibitem{SB18}
D.~Skoupil, P.~Byd\v{z}ovsk\'{y}, Phys. Rev. C \textbf{97}, 025202 (2018).
                  
\bibitem{IMweb}
D.~Skoupil, P.~Byd\v{z}ovsk\'{y}, Prague Isobar Model Web Interface, \url{http://www.ujf.cas.cz/en/departments/department-of-theoretical-physics/isobar-model.html}

\bibitem{RPR}
T.~Corthals, J.~Ryckebusch, and T.~Van Cauteren,  Phys. Rev. C  \textbf{73},  045207 (2006).

\bibitem{RPR07} T.~Corthals, T.~Van Cauteren, J.~Ryckebusch, and D.~G.~Ireland, Phys. Rev. C {\bf 75}, 045204 (2007); 
T.~Corthals, T.~Van Cauteren, P.~Vancraeyveld,  
J.~Ryckebusch, and D.~G.~Ireland, 
Phys. Lett. B {\bf 656}, 186 (2007). 

\bibitem{RPR11} L.~De Cruz, T.~Vrancx, P.~Vancraeyveld, and J.~Ryckebusch, Phys. Rev. Lett. {\bf 108}, 182002 (2012); 
L.~De Cruz, J.~Ryckebusch, T.~Vrancx, P.~Vancraeyveld, 
Phys. Rev. C {\bf 86}, 015212 (2012).

\bibitem{Guidal} M.~Guidal, J.-M.~Laget, and M.~Vanderhaeghen, 
Nucl. Phys. A {\bf 627}, 645 (1997).

\bibitem{HH}
H.~Haberzettl, Xiao-Yun Wang, and Jun He, Phys. Rev. C \textbf{92}, 055503 (2015).

\bibitem{Pascalutsa} V.~Pascalutsa, 
Phys.~Rev.~D \textbf{58}, 096002 (1998).

\bibitem{Vrancx} T.~Vrancx, L.~De Cruz, J.~Ryckebusch, and P.~Vancraeyveld, 
Phys.~Rev.~C \textbf{84}, 045201 (2011).

\bibitem{Collins}
P.~D.~B.~Collins, \textit{''An Introduction to Regge Theory and High Energy Physics,``} Cambridge Monographs on Mathematical Physics (Cambridge University Press, Cambridge, UK, 1977).

\bibitem{Drell}
S.~D.~Drell and T.~D.~Lee, Phys. Rev. D \textbf{5}, 1738 (1972).

\bibitem{Guidal2}
M.~Guidal, J.~M.~Laget, and M.~Vanderhaeghen, Phys. Rev. C \textbf{68}, 058201(R) (2003).

\bibitem{PDG}
 M. Tanabashi {\it et al.} (Particle Data Group), Phys. Rev. D \textbf{98}, 030001 (2018).

\bibitem{DeCruz}
\textcolor{blue}{L.~De Cruz, J.~Ryckebusch,T.~Vrancx, and P.~Vancraeyveld, Phys. Rev. C \textbf{86}, 015212 (2012)}

\bibitem{DeCruz-PhD}
\textcolor{blue}{L.~De Cruz, PhD Thesis, Ghent University, 2011, \url{https://biblio.ugent.be/publication/2063569}.}

\bibitem{Ronchen}
D.~Rönchen \emph{et al.}, Eur. Phys. J. A \textbf{54},  110 (2018).

\bibitem{Minuit}
F.~James and M.~Roos, Comput. Phys. Commun. \textbf{10}, 343 (1975).

\bibitem{CLAS10}
M.~E.~McCracken \emph{et al.}, Phys. Rev. C \textbf{81}, 025201 (2010).

\bibitem{CLAS05}
R.~Bradford \emph{et al.}, Phys. Rev. C \textbf{73}, 035202 (2006).

\bibitem{LEPS18}
S.~H.~Shiu \emph{et al.}, Phys. Rev. C \textbf{97}, 015208 (2018).

\bibitem{LEPS06}
M.~Sumihama \emph{et al.}, Phys. Rev. C \textbf{73}, 035214 (2006).

\bibitem{AS}
R.~A.~Adelseck and B.~Saghai, Phys. Rev. C \textbf{42}, 108 (1990).

\bibitem{Sarantsev}
A.~V.~Sarantsev \emph{et al.}, Eur. Phys. J. A \textbf{25}, 441  (2005).

\bibitem{ByMa07}P. Byd\v{z}ovsk\'y and T. Mart, Phys. Rev. C {\bf 76}, 065202 (2007).

\bibitem{Dey}
B.~Dey and  C.~A.~Meyer, arXiv:1106.0479v1 [hep-ph].

\bibitem{CLAS16}
C.~A.~Paterson \emph{et al.}, Phys. Rev. C \textbf{93}, 065201 (2016).

\bibitem{RPR-12}P. Byd\v{z}ovsk\'y and D. Skoupil, Nucl. Phys. A {\bf 914}, 14 (2013).

\bibitem{Wang}
Xiao-Yun Wang, Jun He, and H. Haberzettl, Phys. Rev. C \textbf{93}, 045204 (2016).

\bibitem{StrangeCalc}
P.~Vancraeyveld, T.~Vrancx, L.~DeCruz, \url{http://rprmodel.ugent.be/calc/}

\bibitem{DSphd}
D.~Skoupil, \emph{"Electromagnetic Production of Kaons,"} PhD Thesis, Czech Technical University, 2016, \url{https://physics.fjfi.cvut.cz/publications/ejcf/DIS_Dalibor_Skoupil.pdf}

\bibitem{Bleck} 
A.~Bleckmann \emph{et al.}, Z.~Phys. \textbf{239}, 1 (1970).

\bibitem{Brown} 
C.~N.~Brown et al, Phys.~Rev.~Lett. \textbf{28}, 1086 (1972).

\bibitem{E94-107} 
P.~Markowitz and A.~Acha, Int.~J.~Mod.~Phys.~E \textbf{19}, 2383 (2010).

\bibitem{CLAS04}
J.~W.~McNabb \emph{et al.}, Phys. Rev. C \textbf{69}, 042201 (2004).

\bibitem{GRAAL07} 
A.~Lleres \emph{et al.}, Eur.~Phys.~J.~A \textbf{31}, 79 (2007).

\bibitem{GRAAL09}
A.~Lleres \emph{et al.}, Eur.~Phys.~J.~A {\bf 39}, 149 (2009).

\bibitem{CLAS-double} 
R.~Bradford \emph{et al.}, Phys.~Rev.~C \textbf{75}, 035205 (2007).

\bibitem{Landay}
J.~Landay, M.~Mai, M.~Doring, H.~Haberzettl, and K.~Nakayama, Phys. Rev. D \textbf{99}, 016001 (2019).

\bibitem{Kamano}
\textcolor{blue}{
H.~Kamano, S.X.~Nakamura, T.S.H. Lee, and T. Sato, Phys. Rev. C \textbf{88}, 035209 (2013).
}

\bibitem{Hunt}
\textcolor{blue}{
B.~C.~Hunt and D.~M.~Manley, Phys. Rev. C \textbf{99}, 055205 (2019).
}

\bibitem{Hunt2}
\textcolor{blue}{
B.~C.~Hunt and D.~M.~Manley, Phys. Rev. C \textbf{99}, 055204 (2019).
}

\bibitem{HuntPhD}
\textcolor{blue}{
B.~C.~Hunt, PhD Thesis, Kent State University, 2017, \url{https://etd.ohiolink.edu/pg_10?::NO:10:P10_ETD_SUBID:151987#abstract-files}
}

\bibitem{MartSakinah}
\textcolor{blue}{
T.~Mart, S.~Sakinah, Phys. Rev. C \textbf{95}, 045205 (2017)
}

\bibitem{SLAC}
A.~M.~Boyarski \emph{et al.}, Phys. Rev. Lett. 22 (1969) 1131-1133.

\bibitem{SAPHIR}
K.~H.~Glander \emph{et al.}, Eur. Phys. J. A \textbf{19} (2004) 251273.

\end{thebibliography}
\end{document}